\documentclass[amsmath,amssymb,amsbsy,prb,twocolumn,superscriptaddress,showpacs]{revtex4-1}

\usepackage{amsmath}
\usepackage{amssymb}
\usepackage[usenames, dvipsnames]{color} 
\usepackage{dcolumn}
\usepackage[pdftex]{graphicx}
\usepackage{epstopdf}
\usepackage{leftidx}
\usepackage[breaklinks,colorlinks = true,linkcolor = red,urlcolor=cyan,citecolor=red]{hyperref}
\usepackage{mathrsfs}
\usepackage{mathtools}
\usepackage{rotating}
\DeclareMathAlphabet{\mathpzc}{OT1}{pzc}{m}{it}

\begin{document}

\title{$\mathbb{Z}_2$ gauge theory for valence bond solids on the kagome lattice}

\author{Kyusung Hwang}
\affiliation{Department of Physics and Centre for Quantum Materials, University of Toronto, Toronto, Ontario M5S 1A7, Canada}

\author{Yejin Huh}
\affiliation{Department of Physics and Centre for Quantum Materials, University of Toronto, Toronto, Ontario M5S 1A7, Canada}

\author{Yong Baek Kim}
\affiliation{Department of Physics and Centre for Quantum Materials, University of Toronto, Toronto, Ontario M5S 1A7, Canada}
\affiliation{Canadian Institute for Advanced Research/Quantum Materials Program, Toronto, Ontario MSG 1Z8, Canada}
\affiliation{School of Physics, Korea Institute for Advanced Study, Seoul 130-722, Korea}

\date{\today}

\begin{abstract}
We present an effective $\mathbb{Z}_2$ gauge theory that captures various competing phases in spin-1/2 kagome lattice antiferromagnets: the topological $\mathbb{Z}_2$ spin liquid (SL) phase, and the 12-site and 36-site valence bond solid (VBS) phases. 
Our effective theory is a generalization of the recent $\mathbb{Z}_2$ gauge theory proposed for SL phases by Wan and Tchernyshyov.
In particular, we investigate possible VBS phases that arise from vison condensations in the SL.
In addition to the 12-site and 36-site VBS phases, there exists 6-site VBS that is closely related to the symmetry-breaking valence bond modulation patterns observed in the recent density matrix renormalization group simulations.
We find that our results have remarkable consistency with a previous study using a different $\mathbb{Z}_2$ gauge theory. 
Motivated by the lattice geometry in the recently reported vanadium oxyfluoride kagome antiferromagnet, our gauge theory is extended to incorporate lowered symmetry by inequivalent up- and down-triangles.
We investigate effects of this anisotropy on the 12-site, 36-site, and 6-site VBS phases.
The 12-site VBS is stable to anisotropy while the 36-site VBS undergoes severe dimer melting. Interestingly, any analogue of the 6-site VBS is not found in this approach.
We discuss the implications of these findings and also compare the results with a different type of $\mathbb{Z}_2$ gauge theory used in previous studies.
\end{abstract}

\maketitle

\section{Introduction}

\begin{figure*}[t]
 \includegraphics[width=\linewidth]{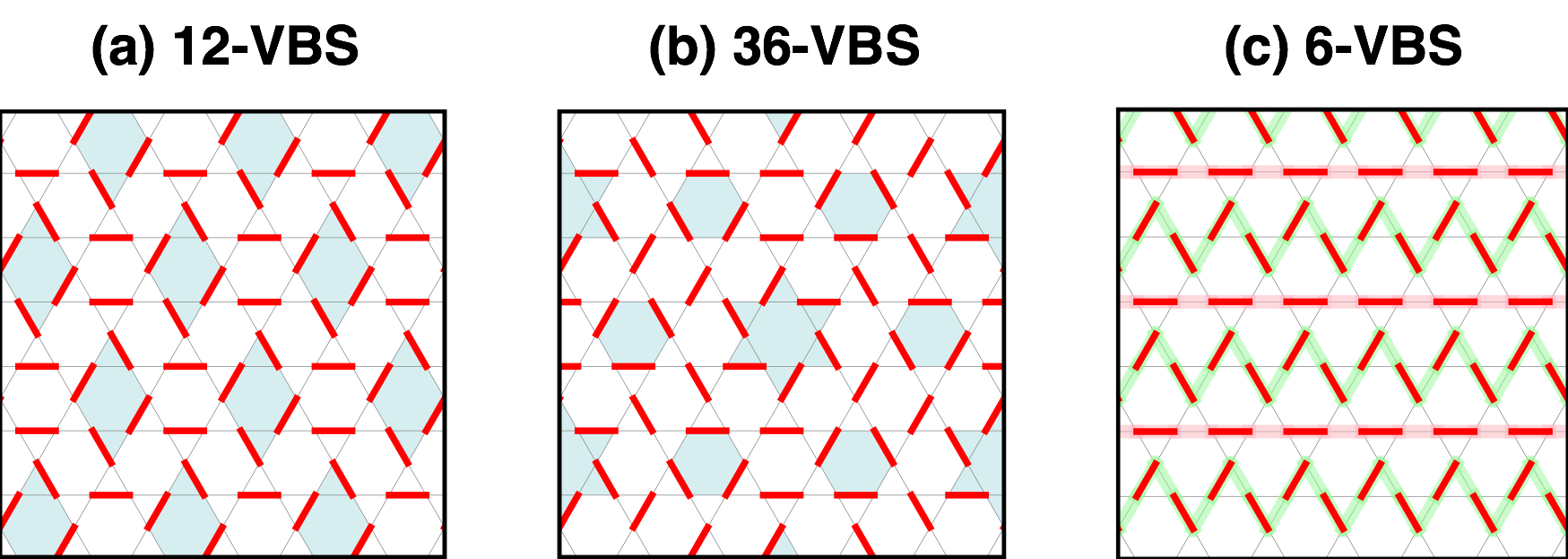}
 \caption{(Color online) Dimer configurations of the 12-site, 36-site, and 6-site VBS phases. (a) The 12-site VBS is characterized by the diamond-shaped dimer patterns (marked with light blue). (b) The 36-site VBS has dimer patterns of the perfect hexagons and stars (marked with light blue). (c) The 6-site VBS features parallel dimers (shaded with pink) and the zigzag dimers (shaded with green). \label{fig:VBSpatterns}}
\end{figure*}

Exploring exotic quantum states of matter has been one of the main themes in condensed matter physics.  In the last two decades, frustrated quantum antiferromagnets have received intense interests due to the capability of harboring various exotic spin states. $\mathbb{Z}_2$ spin liquids (SL) and valence bond solids (VBS) are prominent examples.\cite{Sachdev_2008,Balents_2010} The simplest description of the $\mathbb{Z}_2$ spin liquid can be obtained from a resonating valence bond (RVB) wave function which has characteristics of $\mathbb{Z}_2$ topological order and fractionalized quasiparticle excitations such as spinons and visons.\cite{Anderson_1987,Kivelson_1987,Read_1989,Wen_1991} On the other hand, VBS is a crystalline order of valence bonds, and its elementary excitations are triplons which are confined two-spinon states.

The spin-1/2 nearest-neighbor antiferromagnetic Heisenberg model on the kagome lattice has been extensively studied as a promising model for exotic spin states.
\cite{Zeng_1990, Marston_1991, Leung_1993, Nikolic_2003, Singh_2007, Yang_2008, Evenbly_2010, Sachdev_1992, Wang_2006, Tay_2011, Hastings_2000, Ran_2007, Lu_2011, Iqbal_2011, Iqbal_2011_2, Iqbal_2013, Jiang_2008, Yan_2011, Depenbrock_2012, Huh_2011, Wan_2013, Ju_2013, Elser_1993, Zeng_1995, Mila_1998, Misguich_2002, Poilblanc_2010}
The nature of the ground state in this model system was under debate due to the existence of various competing low energy states. Earlier studies of analytic approaches and numerical computations proposed three major ground state candidates: 
the gapped $\mathbb{Z}_2$ SL,\cite{Sachdev_1992, Wang_2006, Tay_2011, Lu_2011} 
U(1) Dirac SL,\cite{Ran_2007, Iqbal_2011, Iqbal_2011_2, Iqbal_2013} 
and a VBS with a 36-site unit cell\cite{Zeng_1990, Marston_1991, Leung_1993, Nikolic_2003, Singh_2007, Yang_2008, Evenbly_2010, Poilblanc_2010}. 
Recent density matrix renormalization group (DMRG) simulations suggest that this model has a $\mathbb{Z}_2$ SL ground state characterized by a $\mathbb{Z}_2$ topological order and a finite energy gap to excitations.\cite{Yan_2011, Depenbrock_2012} The simulations also reveal diamond-shaped patterns in valence bond correlations of the SL state, indicating that the long range ordered counterpart of the state may be a VBS with a 12-site unit cell.\cite{Huh_2011}
Although the topological $\mathbb{Z}_2$ SL is the ground state of the nearest-neighbor Heisenberg model, other states mentioned above may also be realized by adding small perturbations to the Heisenberg model. For example, the  36-site VBS is stabilized by second nearest-neighbor ferromagnetic interactions as shown in a previous variational Monte Carlo approach.\cite{Iqbal_2011} With many low energy states in competition, the kagome antiferromagnets provide a playground for various competing SL and VBS phases.

On the experimental side, several materials realize the spin-1/2 kagome lattice antiferromagnet.
Herbertsmithite [ZnCu$_3$(OH)$_6$Cl$_2$],\cite{Shores_2005,Mendels_2007,Helton_2007,Imai_2008,Olariu_2008,Han_2012,Pilon_2013, Cepas_2008, Huh_2010, Messio_2010, Potter_2013, Huh_2013, Dodds_2013, Punk_2014}
 the deformed kagome antiferromagnet [Rb$_2$Cu$_3$SnF$_{12}$],\cite{Morita_2008,Matan_2010,Grbic_2013, Yang_2009, Hwang_2014} and the recently discovered vanadium oxyfluoride [(NH$_4$)$_2$(C$_7$H$_{14}$N)(V$_7$O$_6$F$_{18}$)]\cite{Aidoudi_2011,Clark_2013} are such examples.
In these materials, there are generically small perturbations to the nearest-neighbor model on the ideal lattice, such as further neighbor interactions, Dzyaloshinskii-Moriya interactions, and lattice distortions. As an example, vanadium oxyfluoride has a  kagome lattice formed by V$^{4+}$ spin-1/2 moments with inequivalent up- and down-triangles which lowers the sixfold rotation symmetry down to threefold.
Such small perturbations can easily influence the ground state realized in materials with various competing phases. Therefore, it will be useful to understand the competing phases and the transitions between them, based on symmetries that are broken by the small perturbations.

In this work, we develop an effective $\mathbb{Z}_2$ gauge theory, or equivalently a quantum dimer model (QDM), for competing phases in kagome antiferromagnets. The effective $\mathbb{Z}_2$ gauge theory enables us to study competing $\mathbb{Z}_2$ SL and VBS phases on equal footing by considering symmetry-allowed interactions in the corresponding QDM.\cite{Rokhsar_1988,Moessner_2001}
Among previous QDM approaches on kagome antiferromagnets,\cite{Elser_1993,Zeng_1995,Mila_1998,Misguich_2002,Nikolic_2003,Poilblanc_2010,Wan_2013}
our theory is constructed based on the recent $\mathbb{Z}_2$ gauge theory by Wan and Tchernyshyov in Ref. [\onlinecite{Wan_2013}].
They used the effective $\mathbb{Z}_2$ gauge theory to understand the topological $\mathbb{Z}_2$ SL phase and its leading valence bond correlations, identified in the DMRG computations\cite{Yan_2011} on the nearest-neighbor Heisenberg model on the ideal kagome lattice.
Here we generalize this theory to understand possible VBS phases and also extend the theory to incorporate the lowered rotational symmetry motivated by the vanadium oxyfluoride compound.
With this extension, we have another parameter for studying nearby VBS and $\mathbb{Z}_2$ SL phases.

On the ideal kagome lattice, we find that our $\mathbb{Z}_2$ gauge theory captures not only the 12-site VBS but also the 36-site VBS mentioned earlier [see Fig. \ref{fig:VBSpatterns} (a) and (b)]. In addition, we discover another interesting phase, the 6-site VBS [Fig. \ref{fig:VBSpatterns} (c)] which is closely related to a particular feature of the $\mathbb{Z}_2$ SL identified in DMRG calculations, {\it i.e.} symmetry-breaking valence bond modulations on certain cylinder geometries.\cite{Yan_2011,Wan_2013,Ju_2013}
These results show a remarkable consistency with the previous work considering a different form of $\mathbb{Z}_2$ gauge theory for the ideal kagome lattice.\cite{Huh_2011}
Almost all VBS patterns found in the latter are also identified in our $\mathbb{Z}_2$ gauge theory. 
Combining our results on the VBS phases with the results in Ref. [\onlinecite{Wan_2013}] on the $\mathbb{Z}_2$ SL, we show that our gauge theory provides a simple but generic model that captures various competing phases such as the $\mathbb{Z}_2$ SL phase, the 12-site, 36-site, and 6-site VBS phases.

When the lowered symmetry by inequivalent up- and down-triangles is incorporated in the theory, we observe interesting effects of this on the VBS phases found on the ideal kagome lattice. 
First, the 12-site VBS is generally stable to this perturbation. 
In the 36-site VBS, however, we find significant dimer melting, particularly at the ``star" dimer structures while the perfect hexagon patterns remain relatively robust [the star and perfect hexagon dimer structures are marked with light blue in Fig. \ref{fig:VBSpatterns} (b)].
Interestingly, the 6-site VBS does not appear in the lowered symmetry environment.
In the later part of this paper, we will discuss the implications of these findings and also compare the results with a different type of $\mathbb{Z}_2$ gauge theory.

\subsection{Hamiltonian and overview}

Here we introduce our model Hamiltonian and provide an overview of this paper. For an intuitive picture, we start with a quantum dimer model on the kagome lattice that is {\it equivalent} to the $\mathbb{Z}_2$ gauge theory.

The QDM is defined in the Hilbert space of spin-singlet (or dimer) product states. Each state satisfies the so-called hardcore dimer constraint, {\it i.e.} each site on the lattice is covered by only one dimer. In the model, motions and interactions of the dimers are described by the following Hamiltonian:
\begin{eqnarray}
H_{QDM}
&=&
-h \sum_{D\in{\mathscr{D}}} | \bar{D} \rangle \langle D | + K \sum_{D\in{\mathscr{D}}} \epsilon_D | D \rangle \langle D |
\nonumber\\ 
&&
+A \sum_{T\in{\mathscr{T}}} \eta_T | T \rangle \langle T |.
\label{eq:QDM}
\end{eqnarray}
The first term with the coupling $h$ describes dimer motions along various transition graphs (closed loops along which the dimers move) around each hexagon plaquette on the kagome lattice. The 32 transition graphs are listed in Table \ref{tab:trg}. The operator $| \bar{D} \rangle \langle D |$ generates dimer motions between dimer configurations $| D \rangle$ and  $| \bar{D} \rangle$. As an example, Fig. \ref{fig:dimer-motion-int} (a) shows the two dimer configurations on the 12-length transition graph. The set $\mathscr{D}$ consists of all the dimer configurations included in the transition graphs.

The second term with the coupling $K$ represents dimer interactions with the energy coefficient $\epsilon_D$ of the configuration $| D \rangle$. Values of $\epsilon_D$, listed in the last column of Table \ref{tab:trg}, are determined from the $\mathbb{Z}_2$ gauge theory as we will show later. Depending on the sign of $K$, the dimer interactions energetically favor different configurations; for $K>0$, the diamond- and star-shaped dimer configurations are favored, while for $K<0$, the hexagon- and doubled-diamond-shaped configurations are preferred [see Fig. \ref{fig:dimer-motion-int} (b) and (c)]. The Hamiltonian $H_{QDM}$ with $K>0$ and $A=0$ corresponds to the QDM description of the $\mathbb{Z}_2$ gauge theory considered by Wan and Tchernyshyov.

The third term with the coupling $A$ plays the role of a dimer potential energy. This dimer potential term incorporates the lower symmetry as observed in the vanadium oxyfluoride compound by assigning different energy values to the triangles with energy coefficient $\eta_T$ as listed in Table \ref{tab:potential}. The dimer potential energy $\eta_T$ is determined from the $\mathbb{Z}_2$ gauge theory shown later. If $A>0$, dimers tend to occupy the up-triangles while leaving the down-triangles empty and vice versa. Therefore this potential energy term reflects the triangle inequivalence.

\begin{table}
\begin{ruledtabular}
\begin{tabular}{cccc}
Length & Transition graph and $| D \rangle$ & Multiplicity & $\epsilon_D$
\\
\hline
6 & \parbox{5em}{\includegraphics[width=0.6\linewidth]{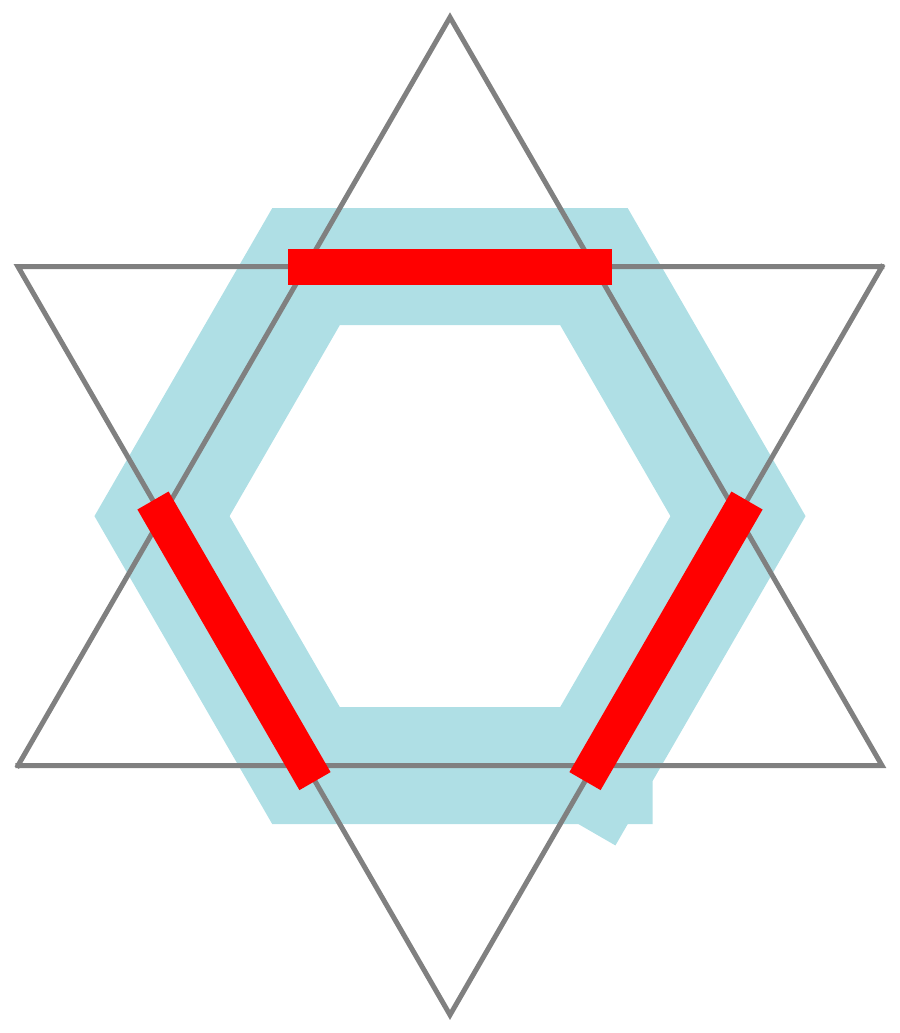}} & 1 & 3
\\
\hline
8 & \parbox{5em}{\includegraphics[width=0.6\linewidth]{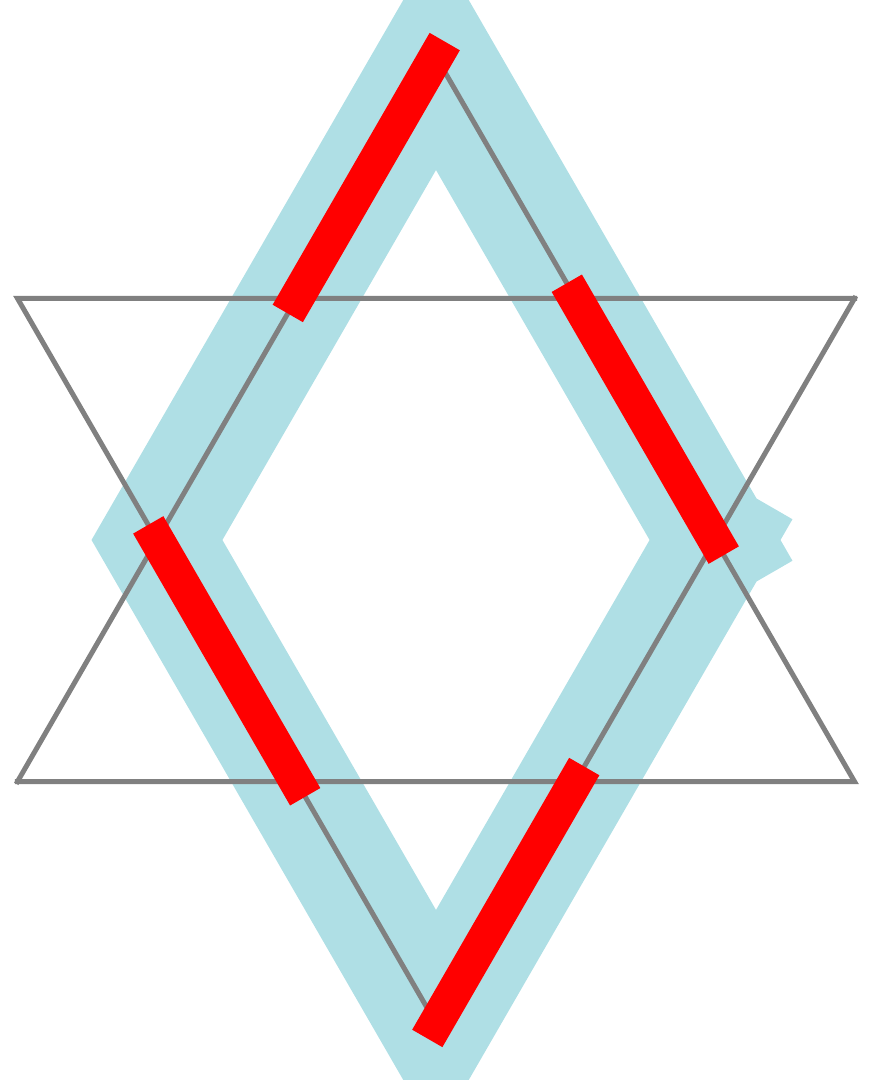}} & 3 & -3
\\
8 & \parbox{5em}{\includegraphics[width=0.6\linewidth]{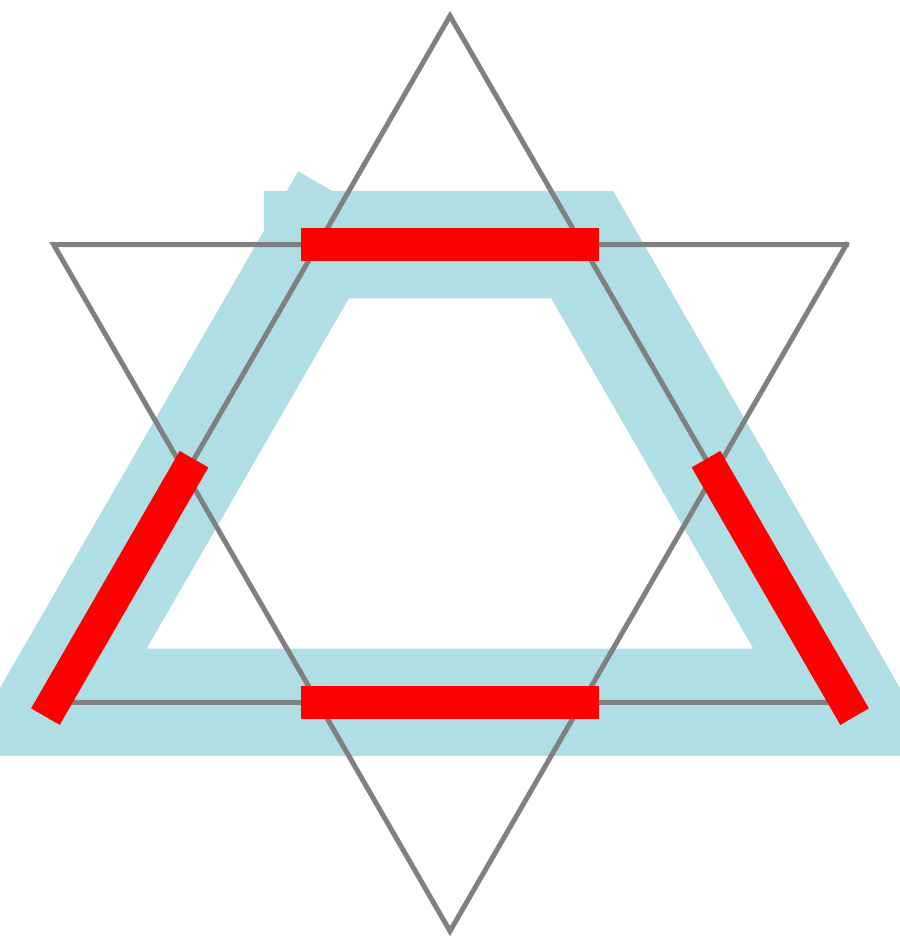}} & 6 & -1
\\
8 & \parbox{5em}{\includegraphics[width=0.6\linewidth]{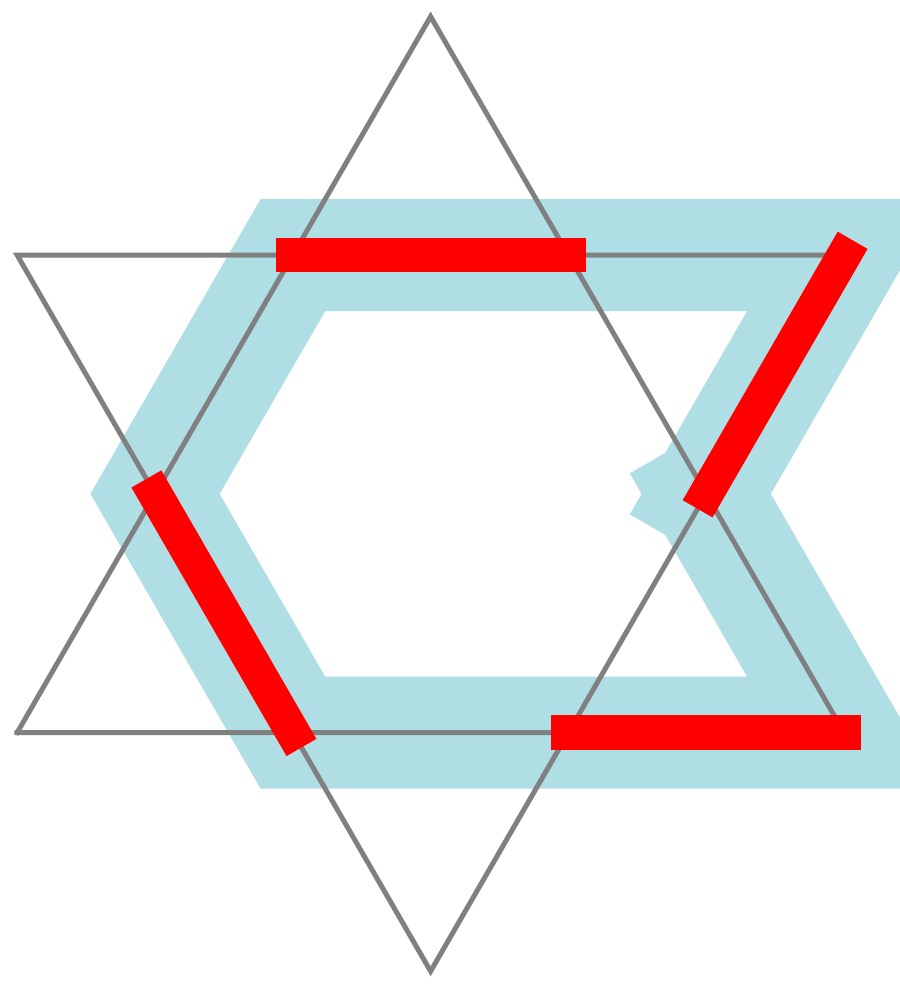}} & 6 & 1
\\
\hline
10 & \parbox{5em}{\includegraphics[width=0.6\linewidth]{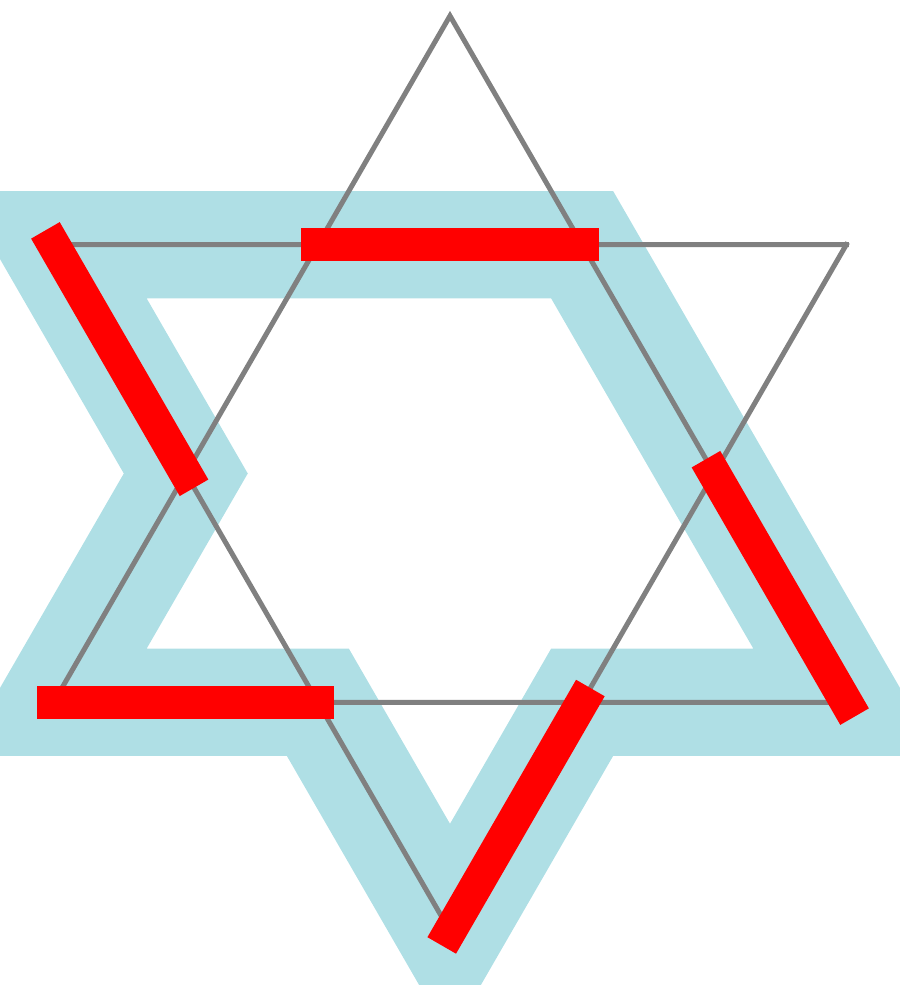}} & 6 & -1
\\
10 & \parbox{5em}{\includegraphics[width=0.6\linewidth]{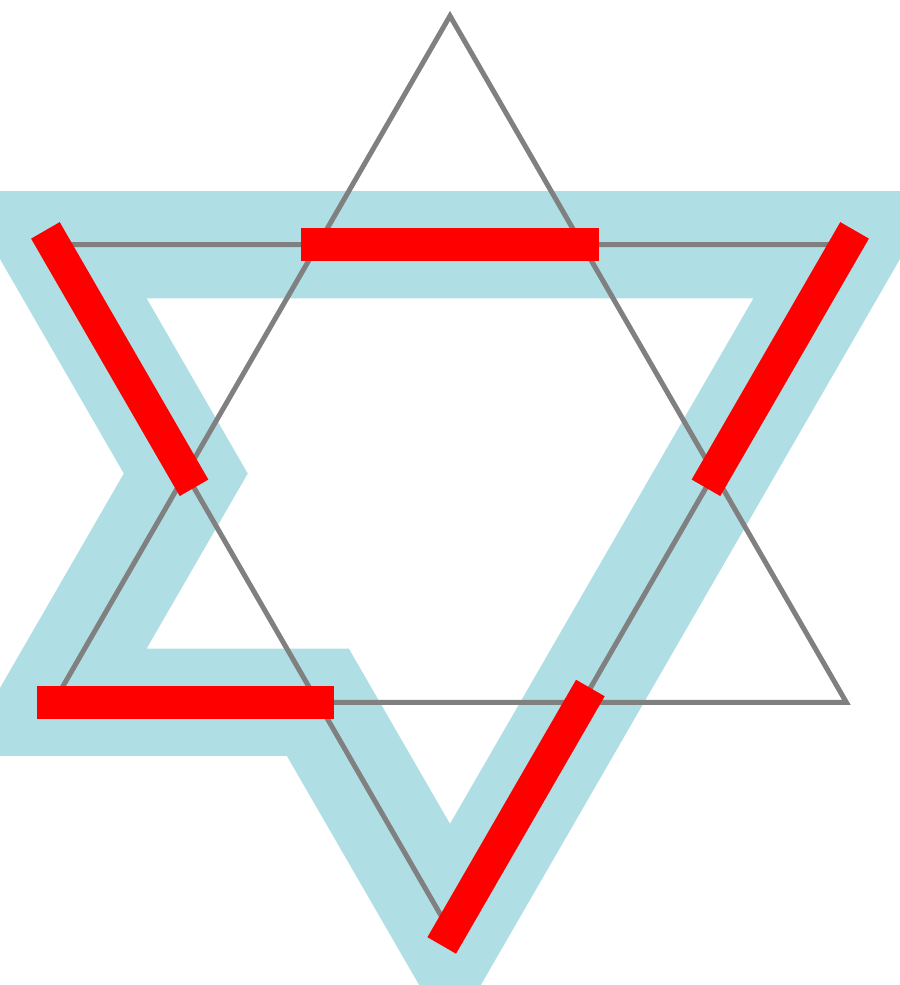}} & 6 & 1
\\
10 & \parbox{5em}{\includegraphics[width=0.6\linewidth]{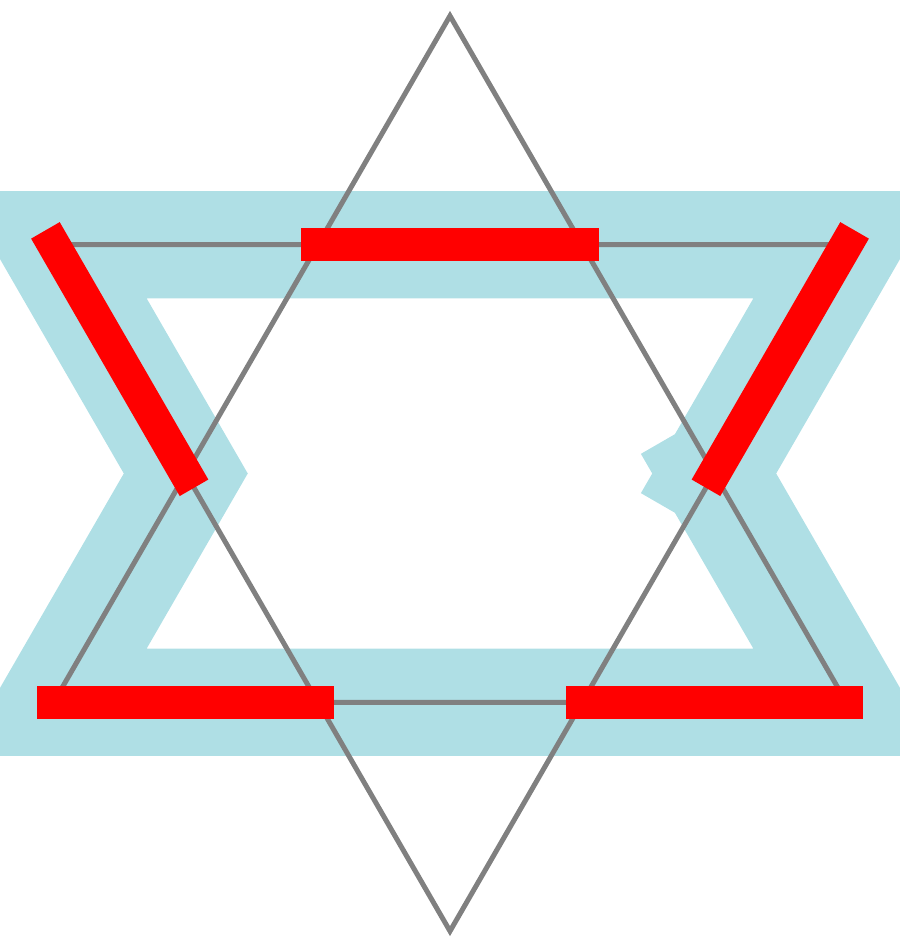}} & 3 & 3
\\
\hline
12 & \parbox{5em}{\includegraphics[width=0.6\linewidth]{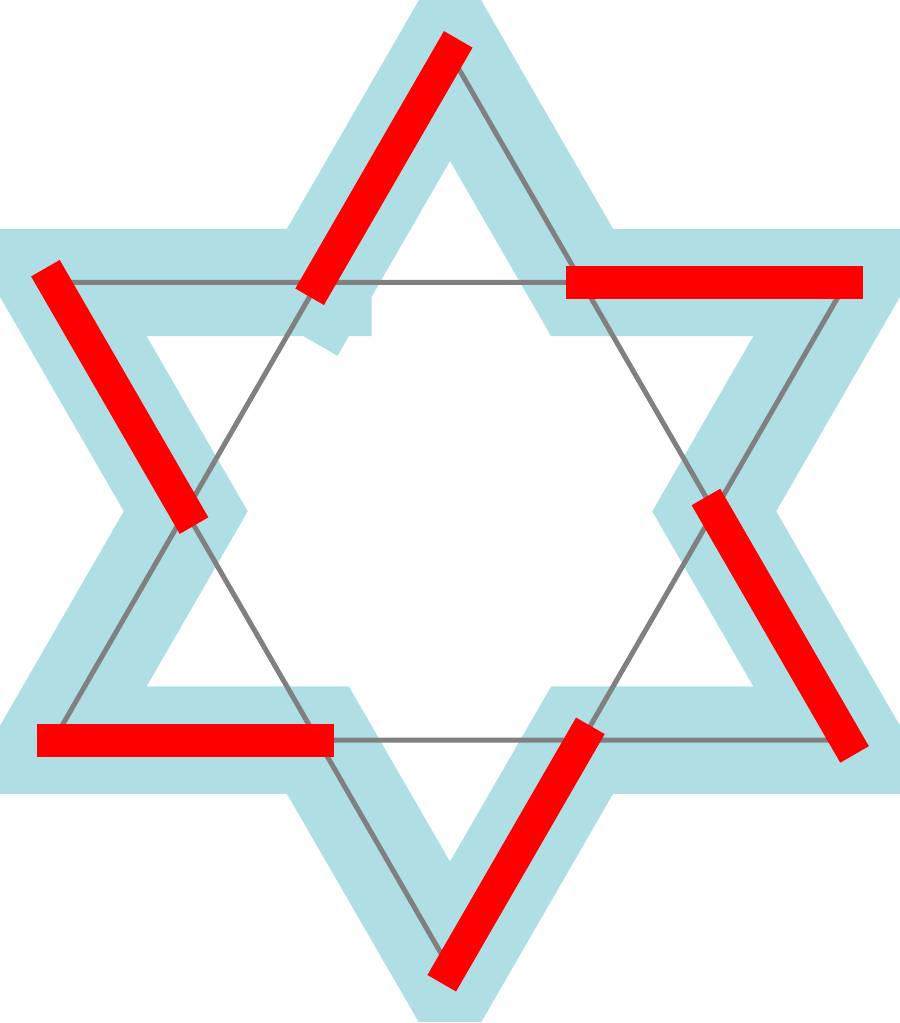}} & 1 & -3
\end{tabular}
\end{ruledtabular}
\caption{(Color online) Transition graphs for the dimer motions and values of the interaction energy coefficient $\epsilon_D$. The transition graphs are denoted with light blue lines. In each graph, one of the dimer configurations ($| D \rangle$) is depicted with red lines. The third column counts the number of equivalent graphs for a given shape.
\label{tab:trg}}
\end{table}

\begin{table}
\begin{ruledtabular}
\begin{tabular}{ccc}
$| T \rangle$ &  & $\eta_T$
\\
\hline
\\
\parbox{3em}{\includegraphics[width=0.6\linewidth]{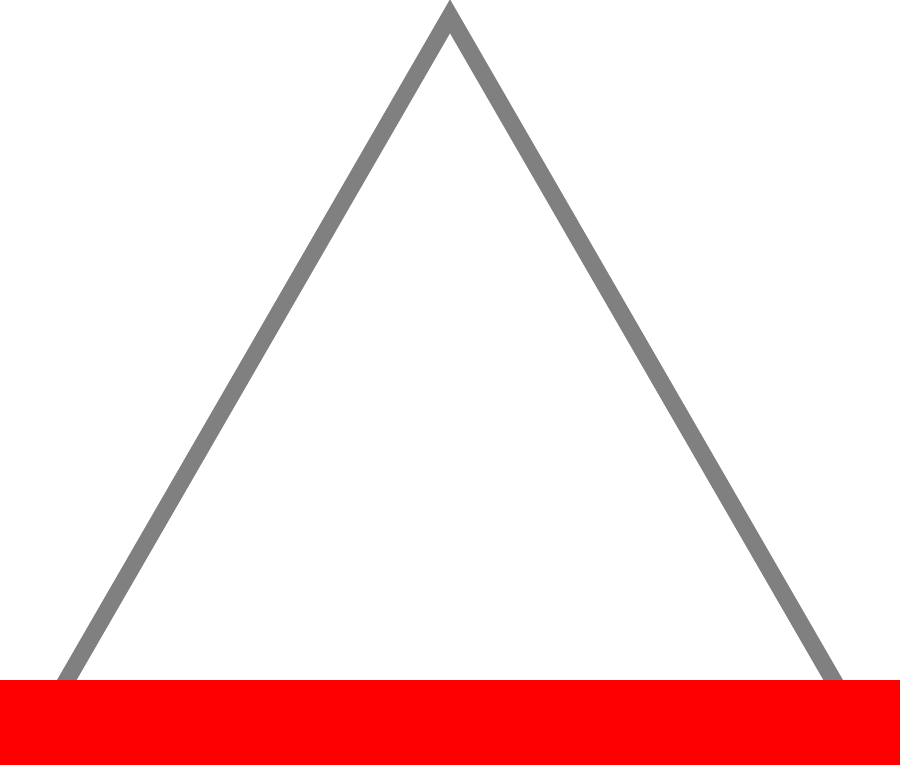}} & & -1/2
\\
\\
\parbox{3em}{\includegraphics[width=0.6\linewidth]{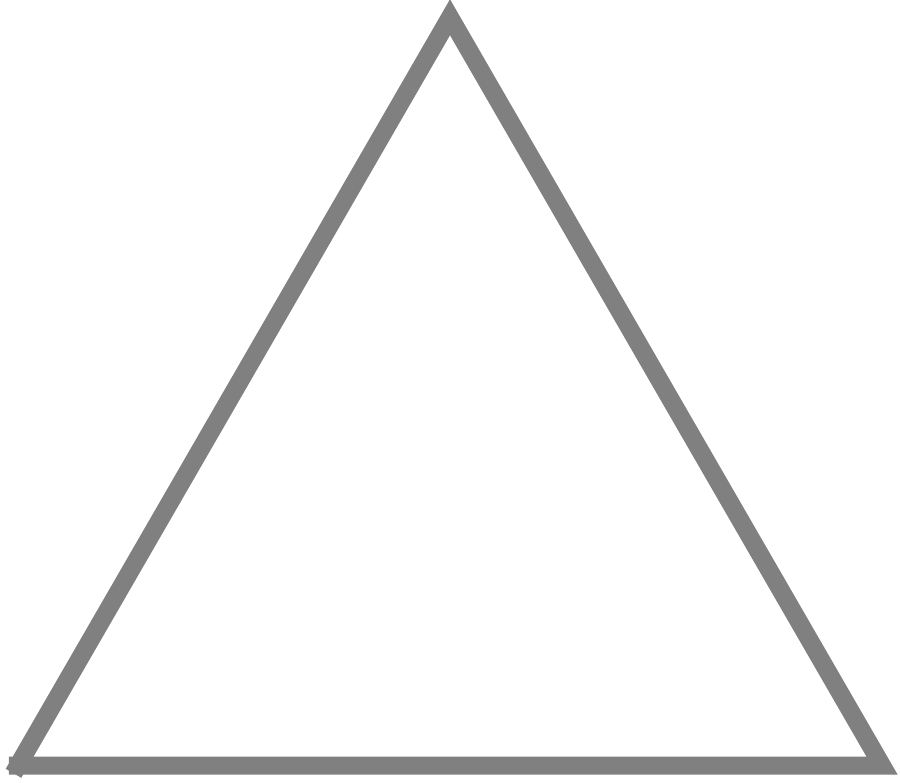}} & & 3/2
\\
\\
\parbox{3em}{\includegraphics[width=0.6\linewidth]{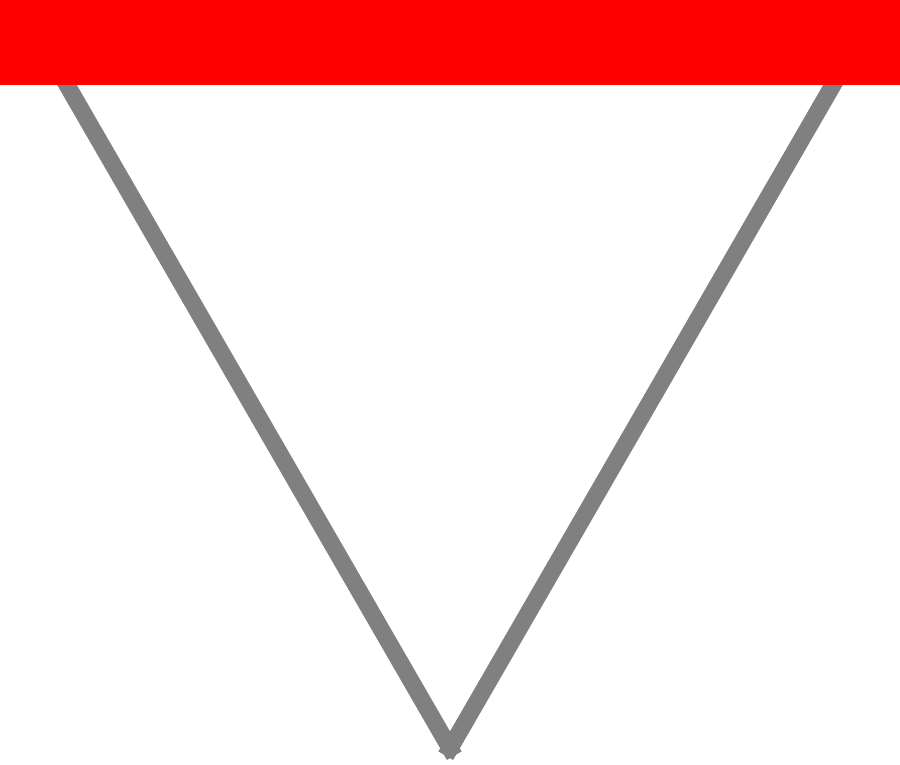}} & & 1/2
\\
\\
\parbox{3em}{\includegraphics[width=0.6\linewidth]{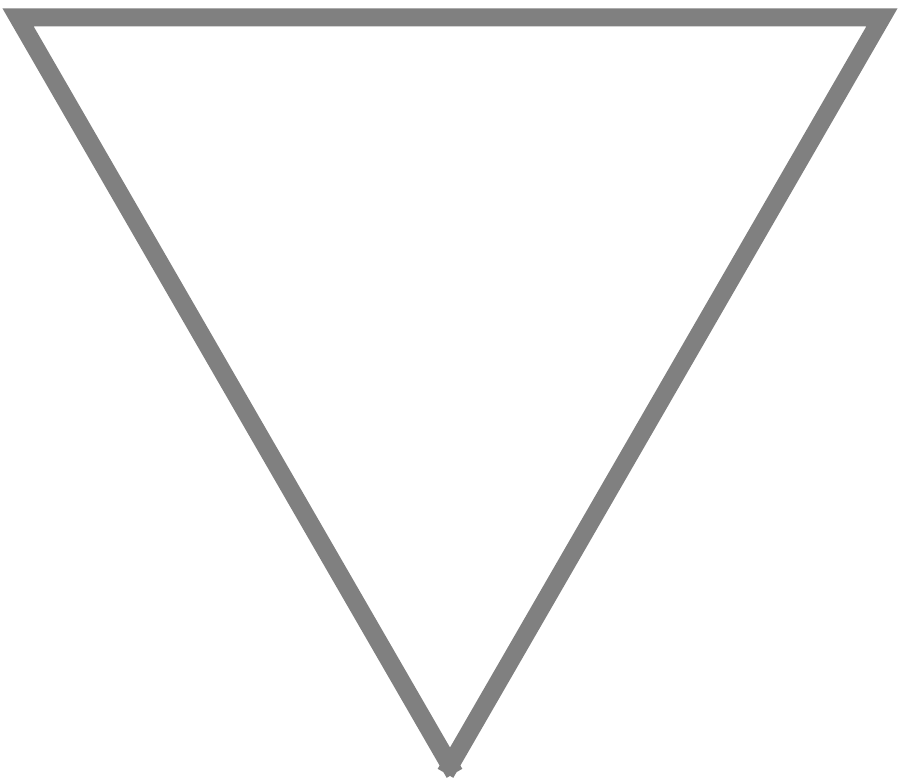}} & & -3/2
\\
\\
\end{tabular}
\end{ruledtabular}
\caption{(Color online) Values of the potential energy coefficient $\eta_T$ for various dimer configurations on up- and down-triangles. Dimers are denoted with red lines.
\label{tab:potential}}
\end{table}

Although it is not so obvious in this form, the QDM (and the equivalent $\mathbb{Z}_2$ gauge theory) possesses two exactly solvable cases: (i) $K=A=0$ and (ii) $h=0$. In the former case, the ground state is exactly expressed by the RVB wave function on the kagome lattice, {\it i.e.} the superposition of all dimer configurations with equal amplitudes, as pointed by Misguich, Serban, and Pasquier in Ref. [\onlinecite{Misguich_2002}]. This special point represents a $\mathbb{Z}_2$ spin liquid phase. In the latter case, the ground state manifold contains various degenerate dimer configurations (singlet product states), each of which is an eigenstate of $H_{QDM}$ and corresponds to a valence bond solid phase. For instance, when $K>0$ and $h=A=0$, we find the singlet product state representing the 12-site VBS [Fig. \ref{fig:VBSpatterns} (a)] in the ground state manifold. When $K<0$ and $h=A=0$, the singlet product states corresponding to the 36-site VBS and 6-site VBS [Fig. \ref{fig:VBSpatterns} (b) and (c)] are found.

\begin{figure}[b]
 \includegraphics[width=0.9\linewidth]{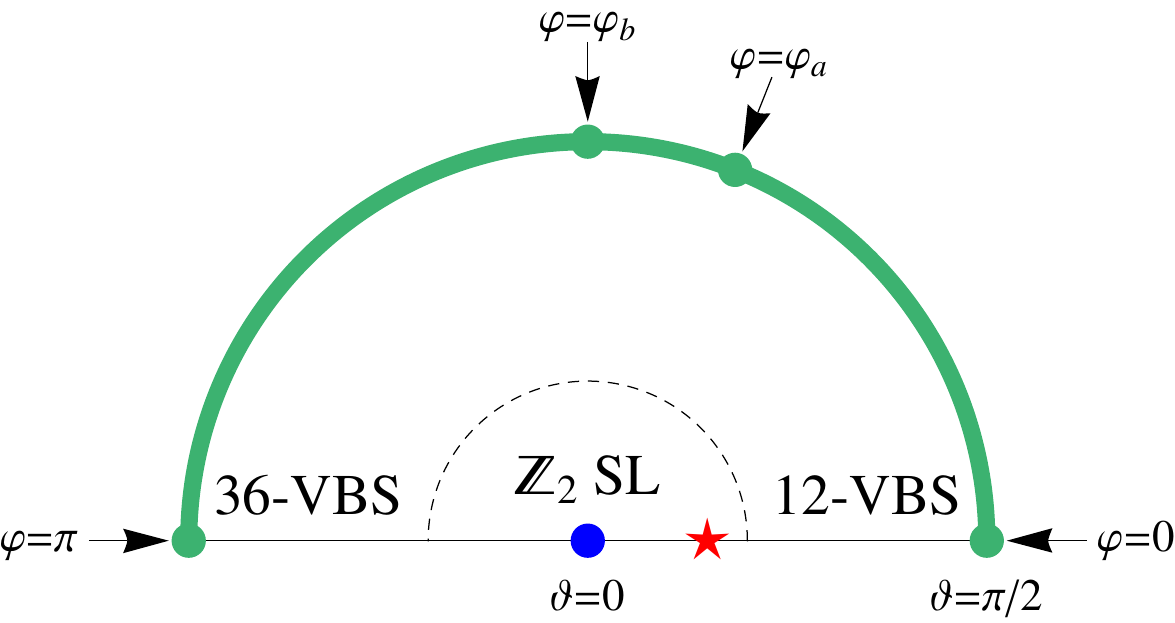}
 \caption{(Color online) Schematic phase diagram of the $\mathbb{Z}_2$ gauge theory. The exactly solvable points are denoted by the blue dot ($K=A=0;~\vartheta=0$) and the green line ($h=0;~\vartheta=\pi/2$). The red star represents the $\mathbb{Z}_2$ SL ground state of the spin-1/2 nearest-neighbor Heisenberg model in the DMRG simulations. $\varphi \rightarrow -\varphi$ takes $A \rightarrow -A$, which is equivalent to switching the up and down triangles. Without loss of generality we only study $\varphi \ge 0$. 
 \label{fig:schematic-PD}}
\end{figure}

In this paper, we explore the intermediate region between the two exactly solvable cases. For convenience, we use the following parametrization for the coupling constants.
\begin{equation}
\begin{array}{ccl}
h &=& \textup{cos} \vartheta,
\\
K &=& \textup{sin} \vartheta ~ \textup{cos} \varphi,
\\
A &=& \textup{sin} \vartheta ~ \textup{sin} \varphi,
\label{eq:parametrization}
\end{array}
\end{equation}
where $0 \leq \vartheta \leq \pi/2$ and $-\pi < \varphi \leq \pi$.
With this parametrization, we draw a schematic phase diagram in Fig. \ref{fig:schematic-PD}. The exactly solvable cases are denoted by a blue dot ($K=A=0;~\vartheta=0$) and a green line ($h=0;~\vartheta=\pi/2$). The dashed line represents a schematic phase boundary separating the $\mathbb{Z}_2$ SL and VBS phases near the exactly solvable points. In the phase diagram, the schematic location of the $\mathbb{Z}_2$ SL ground state of the nearest-neighbor Heisenberg model is represented by a red star. In this study, we explore the VBS regime in the framework of the SL-to-VBS transition. 

\begin{figure}[b]
 \includegraphics[width=0.8\linewidth]{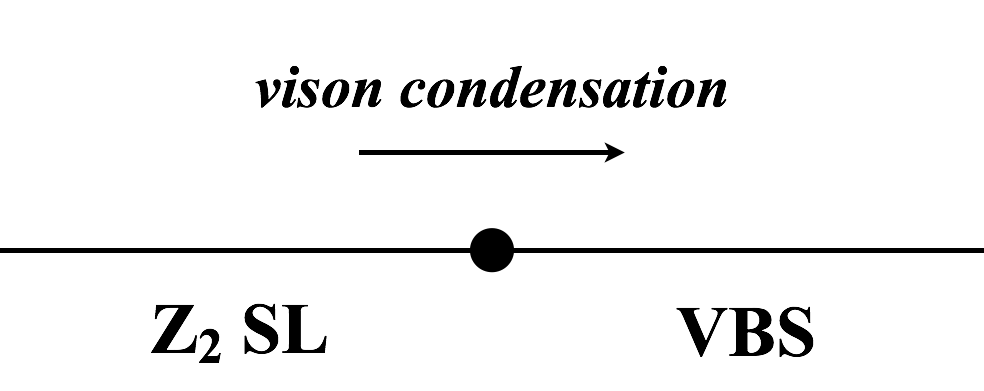}
 \caption{Transition from a $\mathbb{Z}_2$ spin liquid to a valence bond solid by vison condensation.\label{fig:SL-to-VBS}}
\end{figure}

\begin{figure}
 \includegraphics[width=0.8\linewidth]{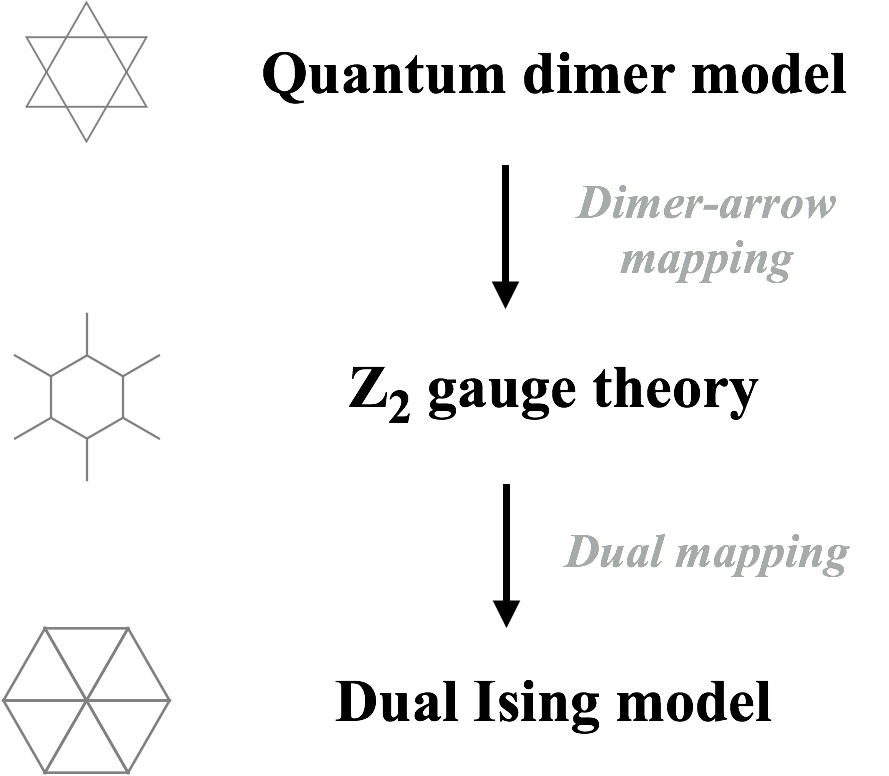}
 \caption{Transformations in the low energy effective theory.\label{fig:trans}}
\end{figure}

In SL-to-VBS transitions, the central objects are the visons, topological vortex excitations in the $\mathbb{Z}_2$ spin liquid phase.\cite{Read_1989,Senthil_2000}
The transition is described by the condensations of visons.\cite{Kogut_1979,Nikolic_2003,Misguich_2008,Xu_2011,Huh_2011} The $\mathbb{Z}_2$ gauge theory provides a convenient platform for the description of the visons. In Sec. \ref{sec:gaugetheory}, we formulate the QDM on the kagome lattice by a $\mathbb{Z}_2$ gauge theory on an effective honeycomb lattice. To describe the visons as local objects, the $\mathbb{Z}_2$ gauge theory is transformed into the dual Ising model on a triangular lattice via a dual mapping in Sec. \ref{sec:dualIsing}. The transformations taken for our low energy theory are summarized in Fig. \ref{fig:trans}. The SL-to-VBS transitions as well as the aforementioned exactly solvable points will be investigated mainly based on the dual Ising model.

In Sec. \ref{sec:landautheory}, we investigate VBS phases appearing from the transitions via Ginzburg-Landau functionals with vison condensation order parameters. The Landau theories are constructed based on projective symmetry group\cite{Wen_2002} (PSG) analysis for the visons and soft mode expansion of the vison field. From the Landau theories and other complementary computations, we find 12-site VBS phases around the $\varphi\simeq0$ line and 36-site and 6-site VBS phases around the $\varphi\simeq\pi$ line in the phase diagram (Fig. \ref{fig:schematic-PD}). Readers who are interested in the resulting VBS phases rather than the computations are advised to jump directly to Sec. \ref{sec:landautheory}.

In Sec. \ref{sec:discussion}, we compare our results with the previous study in Ref. [\onlinecite{Huh_2011}] and discuss influence of the triangle inequivalence, incorporated by the dimer potential energy $A$, on the 12-site, 36-site, and 6-site VBS phases. In addition, we discuss the 6-site VBS as a promising low energy singlet fluctuation in the kagome antiferromagnet and make an interesting remark on it in relation to the recent DMRG simulation results. We provide several computational details in appendices.

\section{$\mathbb{Z}_2$ gauge theory\label{sec:gaugetheory}}

Now we show the explicit construction of our $\mathbb{Z}_2$ gauge theory. Equivalent to the quantum dimer model introduced in the previous section, the gauge theory can be described by a simpler Hamiltonian in which the vison excitations are easily identifiable. We transform the QDM into a gauge theory by employing the dimer-arrow mapping introduced by Elser and Zeng.\cite{Elser_1993} 

\begin{figure}
 \includegraphics[width=0.8\linewidth]{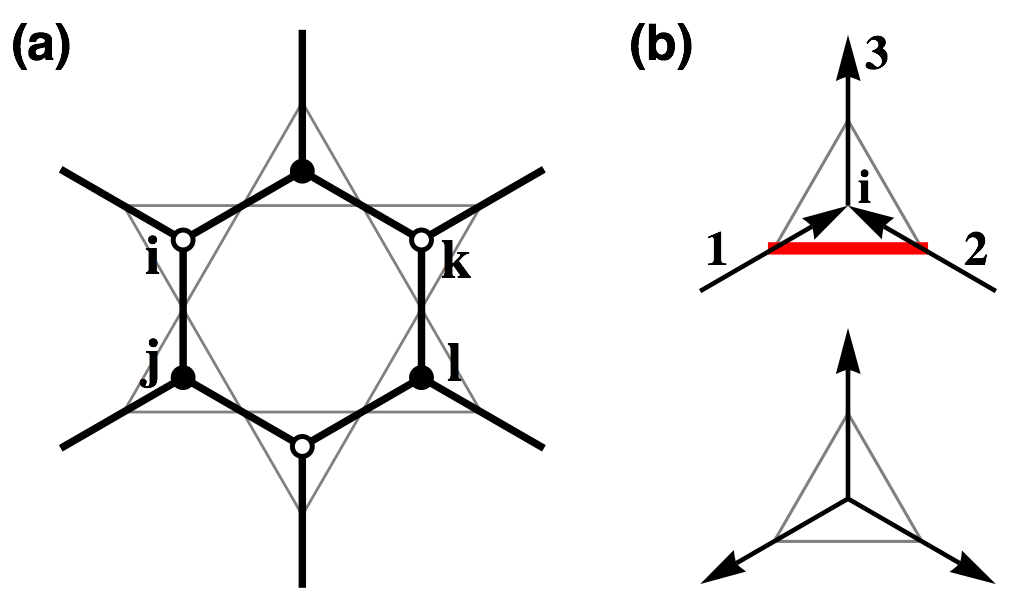}
 \caption{(Color online) Dimer-arrow mapping for the $\mathbb{Z}_2$ gauge theory formulation. (a) The honeycomb lattice, where the arrow representations are defined, is drawn in thick lines. The two sublattices, A and B, are denoted by solid and empty dots. (b) The arrow representations for a dimer-occupied triangle (upper) and an empty triangle (lower). A dimer is denoted by a red line.\label{fig:arrow}}
\end{figure}

Dimers on the kagome lattice are mapped to arrow representations defined on the honeycomb lattice obtained by connecting the centers of the kagome triangles [see Fig. \ref{fig:arrow} (a)]. The dimer-arrow mapping rules are depicted in Fig. \ref{fig:arrow} (b). The two-in-one-out arrow configuration corresponds to a dimer on the kagome link with the two incoming arrows (upper figure). On the other hand, the all-out arrow configuration means there is no dimer on a given kagome triangle (lower figure). By the mapping rules, we establish a one-to-one correspondence between the dimer configurations satisfying the hardcore dimer constraint and the arrow representations.

This can also be expressed by the $\mathbb{Z}_2$ variable $\sigma^x~(=\pm1)$, assigned to each link on the honeycomb lattice. If the arrow on a link $ij$ goes from the sublattice A to B (B to A), we set $\sigma_{ij}^x=+1$ ($\sigma_{ij}^x=-1$). With this $\mathbb{Z}_2$ representation, the hardcore dimer constraint takes the following form.
\begin{equation}
Q_i = \sigma_{i1}^x \sigma_{i2}^x \sigma_{i3}^x
=\left\{
\begin{array}{cc}
+1 & (i \in A)
\\
-1 & (i \in B)
\end{array}
\right.
,
\label{eq:Gauss-law}
\end{equation}
where the sites, 1, 2, 3, represent the three nearest-neighbors of a given site $i$ on the honeycomb lattice [Fig. \ref{fig:arrow} (b)].

\begin{figure}
 \includegraphics[width=0.8\linewidth]{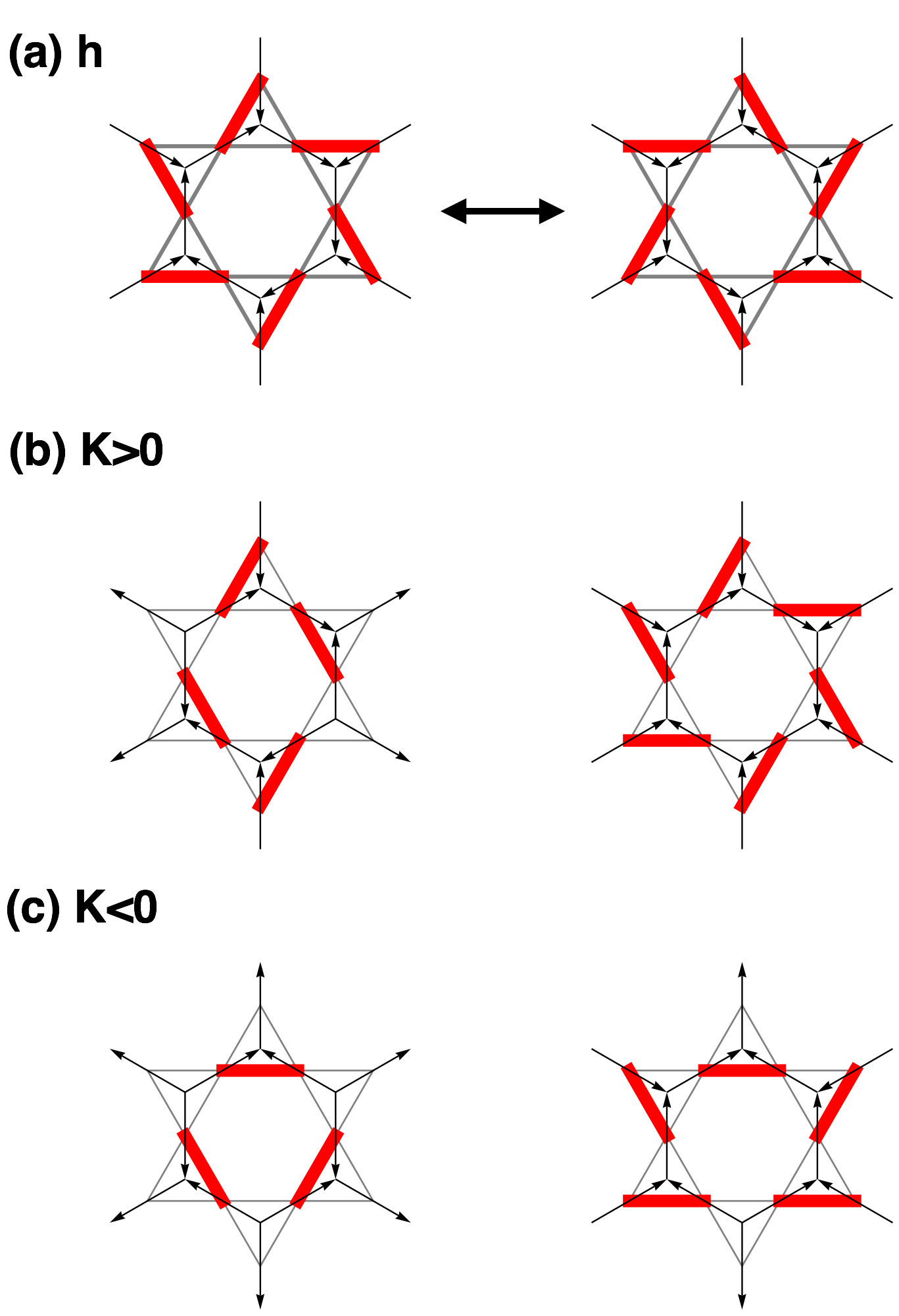}
 \caption{(Color online) Descriptions of the $h$ and $K$ terms in the Hamiltonians, Eqs. (\ref{eq:QDM}) and (\ref{eq:Hamiltonian}). (a) One example of the dimer resonance motions (along the 12-length transition graph in Table \ref{tab:trg}) by the $h$ term. (b) The diamond- and star-shaped dimer patterns are favored by the positive $K$ dimer interactions. (c) The hexagon- and doubled-diamond-shaped dimer patterns are favored by the negative $K$ dimer interactions.
 \label{fig:dimer-motion-int}}
\end{figure}

Dynamics of the dimers can be incorporated by introducing $\mathbb{Z}_2$ variables $\{\sigma_{ij}^z\}$ conjugate to $\{\sigma_{ij}^x\}$. Both variables have Pauli matrix representations as their notations imply. Kinetic terms encoded in the QDM [Eq. (\ref{eq:QDM})] can be represented by a product of $\sigma^z$ variables around a honeycomb plaquette $\alpha$:
\begin{equation}
F_{\alpha}^z=\prod_{\langle ij \rangle \in\alpha} \sigma_{ij}^z .
\end{equation}
This plaquette term encapsulates all the dimer motions described by the 32 transition graphs in Table \ref{tab:trg} [for an example, see Fig. \ref{fig:dimer-motion-int} (a)]. Notice that  $[ F_{\alpha}^z, Q_i]=0$ as the dimer motions respect the hardcore dimer constraint.

In the $\mathbb{Z}_2$ gauge theory description, the QDM takes the following form.
\begin{equation}
 H_{ZGT} = -h \sum_{\alpha} F_{\alpha}^z + K \sum_{\langle ij \rangle// \langle kl \rangle} \sigma_{ij}^x \sigma_{kl}^x + A \sum_{\langle ij \rangle} \sigma_{ij}^x .
 \label{eq:Hamiltonian}
\end{equation}
The first sum represents the dimer motions, and the second and third sums respectively correspond to the dimer interaction and potential energies in the QDM. In the second sum ($K$), each bilinear of $\sigma^x$ is defined on parallel links, $\langle ij \rangle // \langle kl \rangle$, of a honeycomb plaquette as depicted in Fig. \ref{fig:arrow} (a). It is straightforward to check that the sum of the bilinears is equivalent to the dimer interactions in the QDM [some cases in Table \ref{tab:trg} are shown in Fig. \ref{fig:dimer-motion-int} (b) and (c)]. It is also easy to check that the third sum ($A$) represents the dimer potential energy, {\it e.g.}, from Fig. \ref{fig:arrow} (b). We note that the Hamiltonian $H_{ZGT}$ with $K>0$ and $A=0$ is the phenomenological model for the $\mathbb{Z}_2$ spin liquid studied in Ref. [\onlinecite{Wan_2013}]

In this theory, we define the dimer occupation number $d$ on the kagome lattice in the following way.
\begin{equation}
d_{12}^i=\frac{1}{4} (\sigma_{i3}^x - \sigma_{i1}^x - \sigma_{i2}^x ) Q_i + \frac{1}{4}
=\left\{
\begin{array}{cc}
1 & (\textup{occupied})
\\
0 & (\textup{empty})
\end{array}
\right.
.
\label{eq:dimer-occ}
\end{equation}
Here, the dimer occupation number $d_{12}^i$ is defined on the kagome link that the honeycomb links $i1$ and $i2$ cross [see Fig. \ref{fig:arrow} (b)]. It can be easily checked that the above definition applies to both sublattices ($i \in A ~\textup{or}~ B$). Note that the average dimer occupation is $\bar{d}=1/4$ for any hardcore dimer configuration.

Now we interpret the Hamiltonian $H_{ZGT}$ in the language of the $\mathbb{Z}_2$ gauge theory. The Hamiltonian has the $\mathbb{Z}_2$ gauge fields $\{\sigma_{ij}^z\}$ and conjugate electric fields $\{\sigma_{ij}^x\}$ as underlying degrees of freedom. It is written in terms of two gauge-invariant operators, $F_{\alpha}^z$ and $\sigma_{ij}^x$. These operators are invariant under the $\mathbb{Z}_2$ gauge transformation: $\sigma_{ij}^z \rightarrow s_{i} \sigma_{ij}^z s_{j}$ with $s_{i,j}=\pm 1$. The operator $F_{\alpha}^z$ measures the $\mathbb{Z}_2$ magnetic flux through a honeycomb plaquette, {\it e.g.}, 0 flux for $F_{\alpha}^z=1$ and $\pi$ flux for $F_{\alpha}^z=-1$. The electric fields are generated by the Gauss law constraint, Eq. (\ref{eq:Gauss-law}), with the alternating background $\mathbb{Z}_2$ charges $\{Q_i\}$ on the honeycomb lattice.

It is well known that the $\mathbb{Z}_2$ gauge theory has two distinct phases: the deconfined phase when $h \gg |K|,|A|$ and the confined phase when $h \ll |K|,|A|$.\cite{Kogut_1979} In the deconfined phase, the ground state is characterized by a uniform 0 flux and its elementary excitations are the gapped magnetic excitations carrying a $\pi$ flux called the {\it visons}.\cite{Senthil_2000} As depicted in Fig. \ref{fig:dual-lattice} (a), the visons can be created by the string operator 
\begin{equation}
\prod_{\langle ij \rangle \in s} \sigma_{ij}^x,
\label{eq:stringop}
\end{equation}
where the string $s$ extends from the vison position to infinity. The visons are obviously nonlocal objects in the gauge theory. As the electric terms ($K$ and $A$) dominate the magnetic term ($h$), the vison energy spectra get more dispersive due to vison hoppings induced by the electric terms. When the vison gap is closed, vison condensation leads to the confined phase accompanied by lattice symmetry breaking. In terms of the spin model, the deconfined and confined phases correspond to the $\mathbb{Z}_2$ spin liquid and valence bond solid phases, respectively.

\section{Dual Ising model\label{sec:dualIsing}}

In this section, we map the $\mathbb{Z}_2$ gauge theory to the dual Ising model in which the visons become local objects. The dual Ising model enables us to identify the exactly solvable points of the theory and calculate the vison dispersions via a soft spin approximation. We consider four cases that represent different regions of the phase diagram (Fig. \ref{fig:schematic-PD}). For each case, we construct a Ginzburg-Landau functional of the vison fields in Sec. \ref{sec:landautheory}, from the projective symmetry group analysis of the visons.

\begin{figure}
\includegraphics[width=0.8\linewidth]{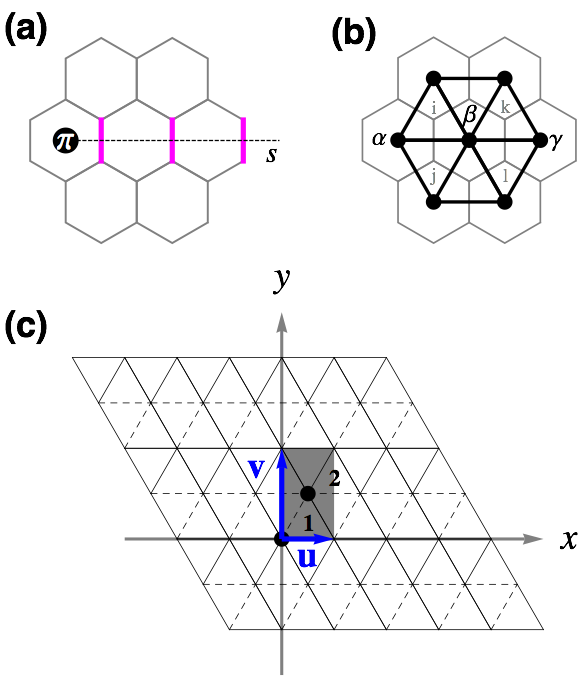}
 \caption{(Color online) (a) Pictorial description of a vison. (b) The dual triangular lattice is depicted in thick lines and its sites are labeled with Greek indices. (c) The gauge choice for $\{ \lambda_{\alpha\beta}\}$ on the dual triangular lattice. The solid and dashed lines represent the links with $\lambda_{\alpha\beta}=+1$ and $\lambda_{\alpha\beta}=-1$, respectively. The shaded rectangle indicates a unit cell in the gauge choice with two sublattices, 1 and 2. The blue arrows are the lattice vectors: ${\bf u}=(1,0)$ and ${\bf v}=(0,\sqrt{3})$.
 \label{fig:dual-lattice}}
\end{figure}

The $\mathbb{Z}_2$ gauge theory on the honeycomb lattice is now transformed into the dual Ising model on a triangular lattice via the following dual mapping.
\begin{equation}
\begin{array}{l}
F_{\alpha}^z=\tau_{\alpha}^x,
\\
\sigma_{ij}^x=\lambda_{\alpha\beta} \tau_{\alpha}^z  \tau_{\beta}^z.
\end{array}
\label{eq:dual-mapping}
\end{equation}
In this mapping, the $\sigma$-operators on the honeycomb lattice are expressed in terms of the new $\tau$-operators on the dual triangular lattice [thick lines in Fig. \ref{fig:dual-lattice} (b)]. The $\tau$-operators satisfy the same Pauli matrix algebra as the $\sigma$-operators. Notice that we are using Latin indices for the honeycomb lattice and Greek indices for the triangular lattice. In this dual description, $\tau^x$ is the vison density operator and $\tau^z$ is the vison creation/annihilation operator. Note that $\tau^z$ and the string operator in Eq. (\ref{eq:stringop}) are essentially the same up to a $\pm$ sign, and the visons are local entities on the dual triangular lattice.
Dynamics of the visons are now governed by the dual Ising model:
\begin{equation}
 H_{DIM} = -h \sum_{\alpha} \tau_{\alpha}^x + K \sum_{\langle\langle\langle \alpha, \gamma \rangle\rangle\rangle}  \lambda_{\alpha\beta} \lambda_{\beta\gamma} \tau_{\alpha}^z \tau_{\gamma}^z + A \sum_{\langle \alpha, \beta \rangle} \lambda_{\alpha\beta} \tau_{\alpha}^z \tau_{\beta}^z .
 \label{eq:dual_Ising}
\end{equation}
Here, the $K$ and $A$ terms provide vison hoppings to the third and first nearest-neighbors on the triangular lattice, respectively.

The additional variable $\lambda~(=\pm 1)$ in the mapping is employed to keep the Gauss law constraint Eq. (\ref{eq:Gauss-law}), which now takes the following form.
\begin{equation}
Q_i = \lambda_{\alpha\beta} \lambda_{\beta\gamma} \lambda_{\gamma\alpha}
=\left\{
\begin{array}{cc}
-1 & (\alpha\beta\gamma = \bigtriangleup)
\\
+1 & (\alpha\beta\gamma = \bigtriangledown)
\end{array}
\right.
.
\label{eq:Gauss-law-2}
\end{equation}
The constraint basically tells us the Berry phase that a vison will acquire when it hops around an elementary triangle. It must be noted that the dual mapping is invariant under the $\mathbb{Z}_2$ gauge transformation: 
 \begin{equation}
 \begin{array}{l}
 \tau_{\alpha}^z \rightarrow G_{\alpha} \tau_{\alpha}^z,
 \\
 \lambda_{\alpha\beta} \rightarrow G_{\alpha} G_{\beta} \lambda_{\alpha\beta},
 \end{array}
 \label{eq:gauge-transformation}
 \end{equation}
 with $G_{\alpha}=\pm 1$.
This fact allows many choices for the $\lambda$-variables. We choose the gauge depicted in Fig. \ref{fig:dual-lattice} (c), which is the same gauge choice taken in Ref. [\onlinecite{Wan_2013}].

\subsection{Exactly solvable points\label{sec:exact}}

The dual Ising model has two exactly solvable cases: (i) $K=A=0$ in the deconfined regime and (ii) $h=0$ in the confined regime. In the former, the ground state is a paramagnetic state with $\tau^x=1$ at every site. In terms of the QDM, it corresponds to the superposition of all the hardcore dimer states with equal amplitudes, which is the resonating valence bond spin liquid state. In the latter case, the model has a degenerate ground state manifold with various hardcore dimer states, or valence bond solid phases, depending on the angle $\varphi$ [Eq. (\ref{eq:parametrization})]. 
When $(\vartheta,\varphi)=(\pi/2,0)$ the degenerate ground states are all symmetry-related states of the 12-site VBS phase. Therefore the 12-site VBS is the only phase in the ground state manifold. In contrast, when  $(\vartheta,\varphi)=(\pi/2,\pi)$ the ground state manifold contains various distinct VBS phases in addition to the  36-site VBS. The 6-site VBS in Fig. \ref{fig:VBSpatterns} (c) is one such phase.
We find that the 12-site, 36-site, and 6-site VBS phases appear in a finite region of the parameter space $\varphi$: 12-VBS in $0 \leq \varphi < \varphi_a$, and 36- and 6-VBS in $\varphi_b < \varphi \leq \pi$. We determine the phase boundaries, $\varphi_a=1.222$ and $\varphi_b=1.571$, by conducting a full search in a 24-site triangular cluster with periodic boundary conditions (which corresponds to a 72-site kagome cluster). Its results are summarized in Fig. \ref{fig:zerohpoints}, which shows the ground state energy and degeneracy as functions of $\varphi$. More details on the $h=0$ case are provided in Appendix \ref{app:zeroh}.

The phase diagram in Fig. \ref{fig:schematic-PD} shows the exactly solvable points with the blue dot ($K=A=0$) and the green line ($h=0$). The intermediate regime between the two cases will be explored in Sec. \ref{sec:landautheory}. 
We shall find the 12-site, 36-site, and 6-site VBS states as the broken symmetry phases that appear via vison condensation from the $\mathbb{Z}_2$ SL (as we move outwards in the phase diagram in Fig.~\ref{fig:schematic-PD}).

\subsection{Soft spin approximation: vison excitations\label{sec:soft}}

Close to the transition point between SL and VBS, it is reasonable to take the following soft spin approximation:
\begin{subequations}
\begin{equation}
 \tau_{\alpha}^z=\pm 1 \rightarrow \phi_{\alpha} \in \mathbb{R},
\end{equation}
\begin{equation}
 -h \sum_{\alpha} \tau_{\alpha}^x \rightarrow \frac{1}{2} \sum_{\alpha} \left( \pi_{\alpha}^2 + m^2  \phi_{\alpha}^2 \right).
\end{equation}
\end{subequations}
Here, the Ising spin $\tau_{\alpha}^z$ is replaced by a real-valued, coarse-grained vison field $\phi_{\alpha}$, and effects of the transverse field term $\tau_{\alpha}^x$ is described by the conjugate momentum $\pi_{\alpha}$ ($[\phi_{\alpha},\pi_{\beta}]=i\delta_{\alpha\beta}$) and a vison mass $m$. The approximation leads to the soft spin Hamiltonian:
\begin{equation}
H_{soft} = \frac{1}{2} \sum_{\alpha} \left( \pi_{\alpha}^2 + m^2  \phi_{\alpha}^2 \right) + \sum_{\alpha \ne \beta} M_{\alpha\beta} \phi_{\alpha} \phi_{\beta} ,
\label{eq:soft-spin-Hamiltonian}
\end{equation}
where the matrix $M_{\alpha\beta}$ contains the hopping amplitudes from the $K$ and $A$ terms. Diagonalizing the Fourier transformed Hamiltonian, we obtain two bands of the vison dispersions, $\omega_{\pm}({\bf q})$, owing to the two sublattices.
\begin{equation}
\omega_{\pm}({\bf q})
=
\sqrt{m^2+K f({\bf q}) \pm |A| \sqrt{6+f({\bf q})}} ,
\label{eq:vison}
\end{equation}
where $f({\bf q}) = 2 \textup{cos}(2{\bf q}\cdot{\bf u})+4\textup{sin}({\bf q}\cdot{\bf u})\textup{sin}({\bf q}\cdot{\bf v})$.

\begin{table}
 \includegraphics[width=0.45\linewidth]{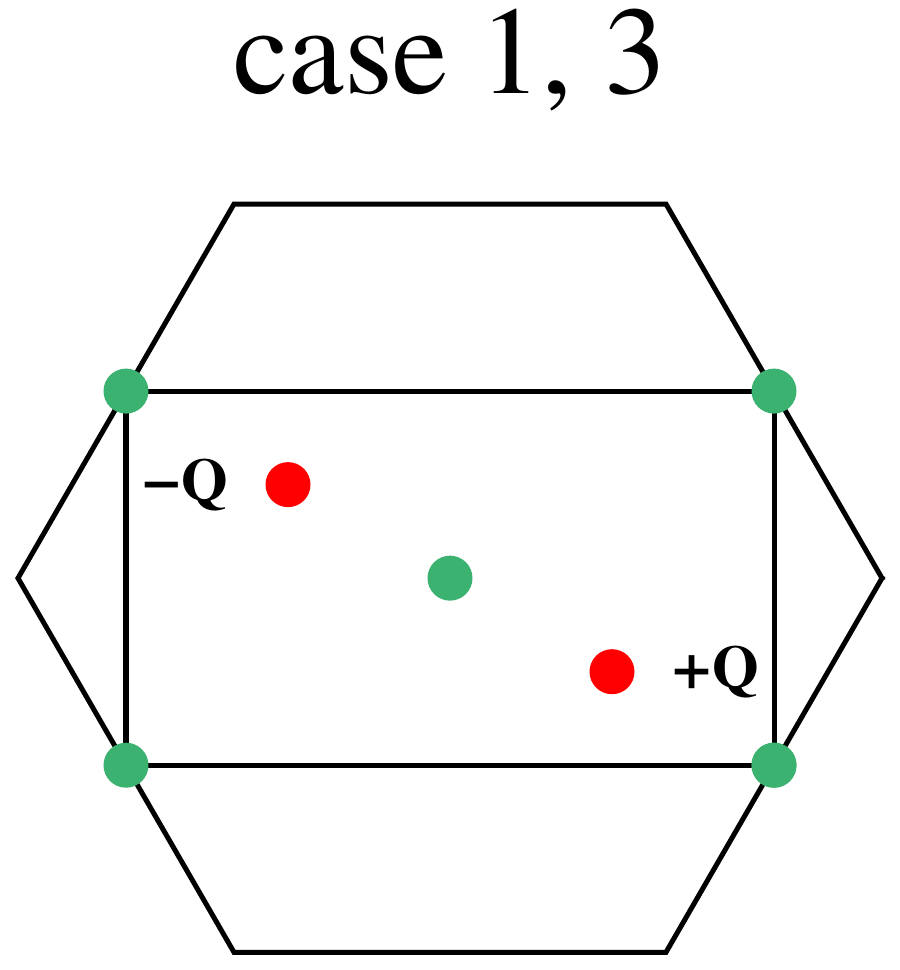}
 ~~~
 \includegraphics[width=0.45\linewidth]{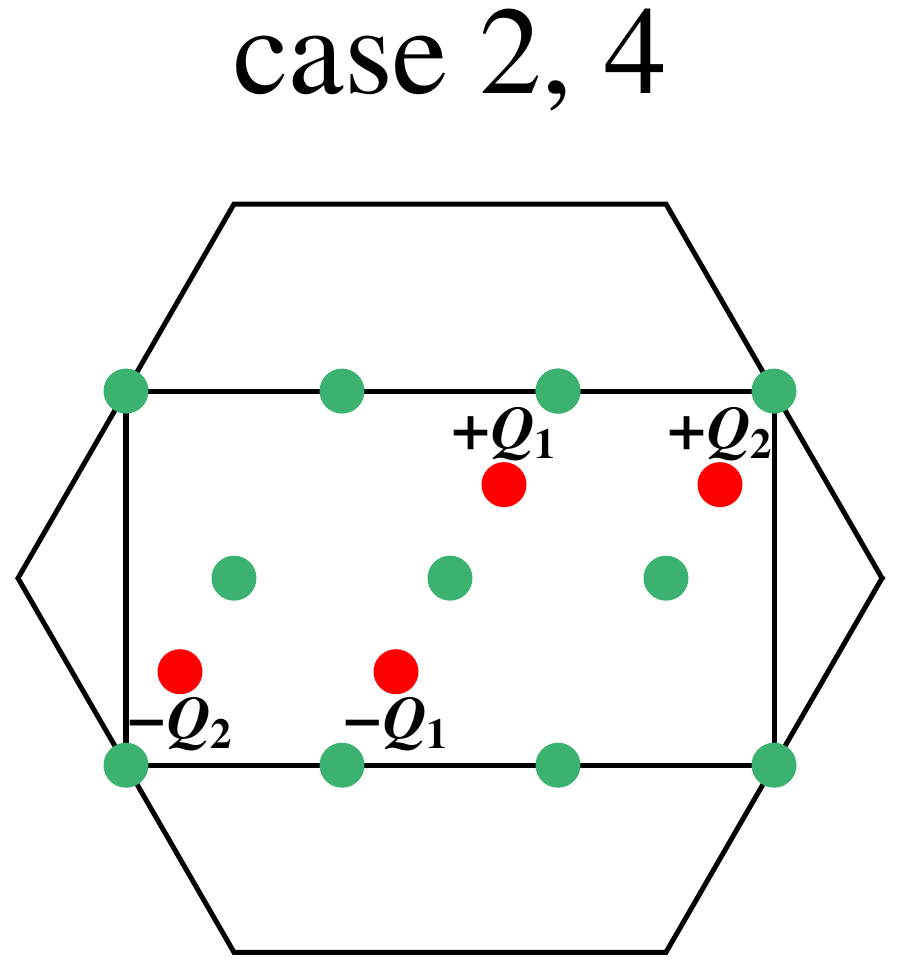}
\\
\begin{ruledtabular}
\begin{tabular}{c|cccc}
case & range & position & degeneracy & $\mathcal{N}$
\\
\hline
1 & $\varphi=0$ & $\pm{\bf Q}~~~$ & 2 & 4
\\
2 & $\varphi=\pi$ & $\pm{\bf Q}_{1,2}$ & 2 & 8
\\
3 & $0< |\varphi| \lesssim \frac{\pi}{3}$ & $\pm{\bf Q}~~~$ & 2 & 4
\\
4 & $\frac{\pi}{2} \lesssim |\varphi| < \pi $ & $\pm{\bf Q}_{1,2}$ & 1 & 4
\end{tabular}
\end{ruledtabular}
\caption{(Color online) Positions and degeneracies of the soft modes in the four cases. The soft mode positions are marked with red dots in the first Brillouin zone (rectangle) of the soft spin Hamiltonian $H_{soft}$; ${\bf Q}=\frac{\pi}{2}(1,-\frac{1}{\sqrt{3}})$, ${\bf Q}_1=\frac{\pi}{6}(1,\sqrt{3})$, and ${\bf Q}_2=\frac{\pi}{6}(5,\sqrt{3})$. The hexagon indicates the Brillouin zone of the kagome lattice. The green dots show positions of the lowest two-vison excitation computed by the equation: $E_2({\bf q}) = \textup{min}_{{\bf k}} [ \omega_-({\bf k}) + \omega_-({\bf q}-{\bf k}) ]$. In the above table, the fourth column, $\mathcal{N}$, is the total number of the soft modes in each case.
}
\label{tab:listofcases}
\end{table}

Positions of the soft modes (the lowest energy modes) in ${\bf q}$-space are marked with red dots in the figures in Table \ref{tab:listofcases}. Depending on the parameters, $K~(\propto\textup{cos}\varphi)$ and $A~(\propto\textup{sin}\varphi)$, the soft modes have different positions and degeneracies as summarized in the table. We consider four cases: (i) $\varphi=0$, (ii) $\varphi=\pi$, (iii) $0< |\varphi| \lesssim \frac{\pi}{3}$, and (iv) $\frac{\pi}{2} \lesssim |\varphi| < \pi $. In cases 1 and 3, four soft modes occur at the points, $\pm{\bf Q}$, with a twofold degeneracy at each point (left figure). Case 2 has eight soft modes at the points, $\pm{\bf Q}_{1,2}$, each twofold degenerate, whereas case 4 has four non-degenerate modes at the same points (right figure). In the remaining $\frac{\pi}{3} \lesssim |\varphi| \lesssim \frac{\pi}{2}$, which is not included in the table, soft modes occur at incommensurate positions, which vary with $\varphi$ and have line degeneracies. Due to this complexity, the last case is not considered in our study. For each case in the table, Landau theory will be developed based on the vison soft modes.

It is important to note that the visons are gauge-dependent objects owing to the gauge freedom in Eq. (\ref{eq:gauge-transformation}). Hence, the vison follows a projective representation of the symmetry group (PSG) of the system and the soft mode positions can change depending on the gauge choice. When periodic boundary conditions are imposed on the system, the visons are always created in pairs by physical operators since $\prod_{\alpha} F_{\alpha}^z=1$. Hence the lowest two-vison excitations are useful in getting physical information about the VBS phase stabilized by the vison condensation. From the lowest modes of the two-vison excitations (green dots in Table \ref{tab:listofcases}), we can figure out the size of the unit cell and the lattice structures of the VBS phases. Specifically, the VBS phases have triangular lattice structures with 12-site unit cells in cases 1 and 3, and also triangular lattice structures but with 36-site unit cells in cases 2 and 4. Detailed VBS structures are studied via Landau theories later.

\subsection{Vison PSG}

\begin{figure*}
 \includegraphics[width=\linewidth]{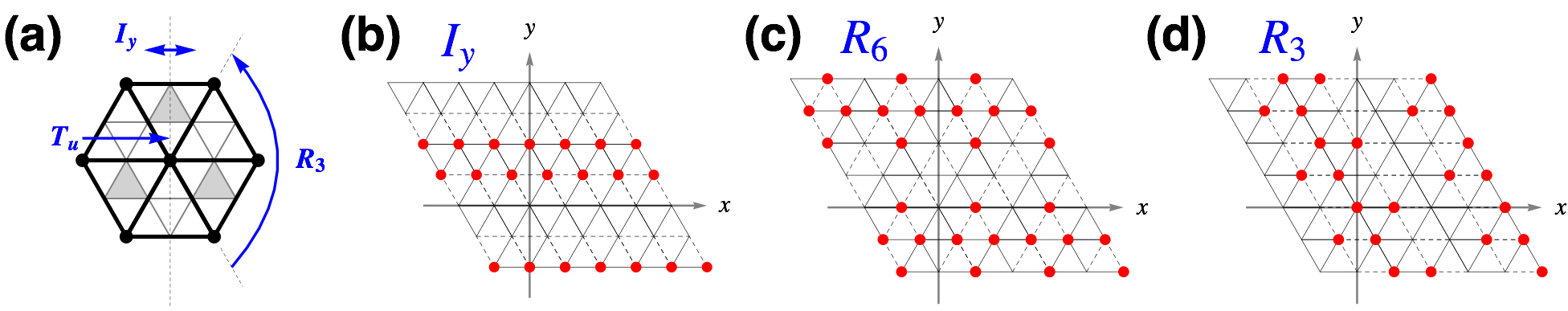}
 \caption{(Color online) Pictorial descriptions of the generators in the vison PSG. 
 (a) The generators of the symmetry group SG$_{3,4}$. $T_u$: translation by the vector ${\bf u}$. $I_y$: reflection with respect to the $y$-axis. $R_3$: rotation by $2\pi/3$ with respect to the origin. For the symmetry group SG$_{1,2}$, the $R_3$-rotation is replaced with the rotation by $\pi/3$ ($R_6$). (b,c,d) The action of the symmetry operation, $X(\in\{I_y, R_6, R_3\})$, on the Hamiltonian $H_{soft}$ with the gauge choice $\{\lambda_{\alpha\beta}\}$ in Fig. \ref{fig:dual-lattice}. Red dots in each figure indicate the sites where the sign changes in the associated gauge transformation $G_X$. The translation operator $T_u$ is not shown here due to its trivial action.
 \label{fig:visonPSG}}
\end{figure*}

Now we analyze the projective symmetry group of the visons in the four cases in Table \ref{tab:listofcases}. Cases 1 and 2 have the space group symmetry SG$_{1,2}$ of the ideal kagome lattice.
\begin{equation}
\textup{SG}_{1,2}=\textup{Span}\{T_u,I_y,R_6\} ,
\end{equation}
where the three generators represent the ${\bf u}$-translation $T_u$, the $y$-axis-reflection $I_y$, and the $\pi/3$-rotation $R_6$ (see Fig. \ref{fig:visonPSG}).
On the other hand, cases 3 and 4 have the symmetry group
\begin{equation}
\textup{SG}_{3,4}=\textup{Span}\{T_u,I_y,R_3\} .
\end{equation}
Here, the rotation symmetry is lowered to the $2\pi/3$-rotation $R_3$ due to the triangle-inequivalence by a nonzero $A$ term. The visons ($\phi_{\alpha}$) transform projectively under the symmetry operation. The projective symmetry group,\cite{Wen_2002} $\textup{PSG}=\{G_X X\}$, is an extended symmetry group with the $\mathbb{Z}_2$ gauge transformation $G_X$ for each symmetry operation $X$ (here, the invariant gauge group is IGG=$\mathbb{Z}_2$). The gauge transformation $G_X$ is determined by the condition that the vison Hamiltonian is invariant under the PSG, {\it i.e.} $M_{X(\alpha)X(\beta)}=M_{\alpha\beta}G_X[X(\alpha)]G_X[X(\beta)]$. We describe the generators of the PSG pictorially in Fig. \ref{fig:visonPSG} and put their explicit expressions in Appendix \ref{app:visonPSG}. 

Next, we turn our attention to the vison condensation order parameters and discuss how to extract a representation of the PSG in the order parameter space. The order parameters are defined by expanding the vison field in soft modes as the following.
\begin{equation}
\phi_s ({\bf r}) = \sum_{n=1}^{\mathcal{N}} \psi_n v_s^n e^{i{\bf q}_n \cdot {\bf r}}.
\label{eq:vison-field}
\end{equation}
Here, ${\bf r}$ and $s~(=1,2)$ are the unit cell and sublattice indices [Fig. \ref{fig:dual-lattice} (c)], and $\mathcal{N}$ is the number of soft modes available. The complex number, $\psi_n$, is the vison order parameter corresponding to the $n$-th soft mode with wave vector ${\bf q}_n$ and eigenvector ${\bf v}^n$ of $H_{soft}$. With the above soft mode expansion, we deduce the PSG of the order parameters as follows.
\begin{eqnarray}
G_X X [ \phi_s({\bf r}) ]
&=& \sum_n \psi_n v_{s'}^n e^{i{\bf q}_n \cdot {\bf r}'} G_X({\bf r}',s')
\nonumber\\
&=& \sum_n {\psi}'_n v_s^n e^{i{\bf q}_n \cdot {\bf r}} .
\end{eqnarray}
In the first line, $({\bf r}',s')=X({\bf r},s)$. The PSG is read off from the second equality in the form of the matrix equation:
\begin{equation}
{\psi}'_m = \sum_n \mathcal{X}_{mn} {\psi}_n .
\end{equation}
The $\mathcal{N}\times\mathcal{N}$ matrix, $\mathcal{X}_{mn}$, is the PSG representation of $X$ in the order parameter space $\{\psi_n\}$. Landau theories in the next section hinge on the above soft mode expansion and the vison PSG.

\section{Landau theories for valence bond solids\label{sec:landautheory}}

We now construct Landau functionals for the four cases in Table \ref{tab:listofcases}. Recall that the four cases represent different regions with different space group symmetries (SG$_{1,2}$ and SG$_{3,4}$) in the phase diagram in Fig. \ref{fig:schematic-PD}.
The Landau theories enable us to investigate possible symmetry breaking patterns that can appear in the valence bond solid phases of the dual Ising model. We discuss the symmetry breaking patterns at mean-field theory level.

\subsection{Case 1: $\varphi=0$}

\begin{figure}[b]
 \includegraphics[width=\linewidth]{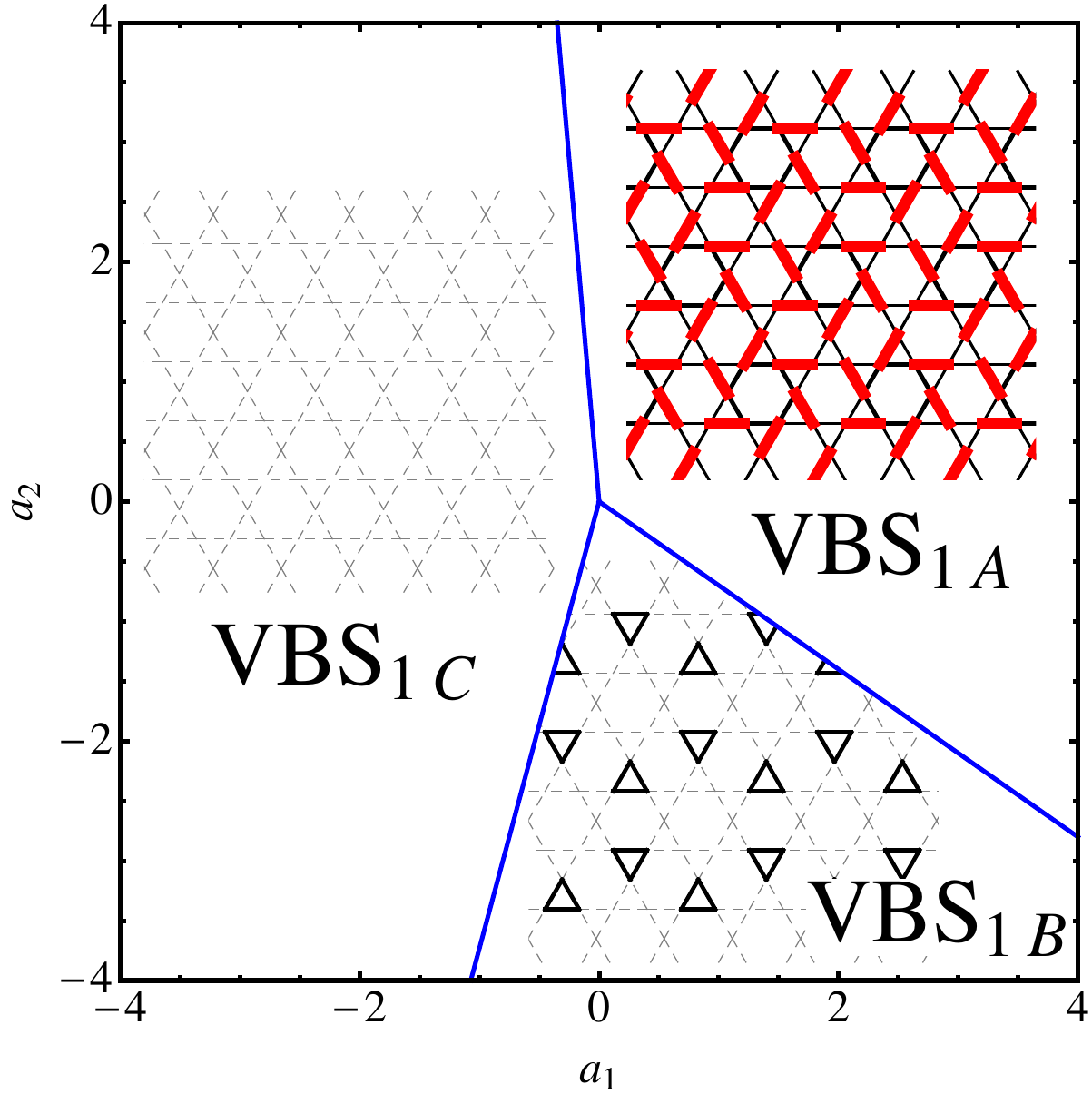}
 \caption{(Color online) Mean-field phase diagram of the Landau theory [Eq. (\ref{eq:case1-GL})] for case 1. VBS$_{1A}$ phase is the 12-site VBS in Fig. \ref{fig:VBSpatterns} (a). The phase diagram is obtained with fixed $u=10$.
 \label{fig:case1-phases}}
\end{figure}

Case 1 has four vison soft modes at $\pm{\bf Q}$ with the symmetry group SG$_{1,2}$. Accordingly, the vison field expansion is 
\begin{equation}
\boldsymbol{\phi} ({\bf r}) = \left( \psi_1 {\bf v}^{(1)} + \psi_2 {\bf v}^{(2)} \right) e^{i{\bf Q}\cdot{\bf r}} + \textup{c.c.}
\label{eq:vison-field-exp}
\end{equation}
with eigenvectors, ${\bf v}^{(1)}=(1,0)^T$ and ${\bf v}^{(2)}=(0,1)^T$, and order parameters, $\Psi=(\psi_1,\psi_2,\psi_1^*,\psi_2^*)$.
The PSG representation carried by the order parameters is generated by the three matrices:
\begin{subequations}
\label{eq:case1-PSG}
\begin{eqnarray}
\mathcal{T}_u
&=&
\left(
\begin{array}{cc|cc}
i&0&0&0
\\
0&i&0&0
\\
\hline
0&0&-i&0
\\
0&0&0&-i
\end{array}
\right),
\\
\mathcal{I}_y
&=&
\left(
\begin{array}{cc|cc}
0&0&1&0
\\
0&0&0&-i
\\
\hline
1&0&0&0
\\
0&i&0&0
\end{array}
\right),
\\
\mathcal{R}_6
&=&
\frac{1}{2}
\left(
\begin{array}{cc|cc}
1&-i&1&-i
\\
-1&-i&1&i
\\
\hline
1&i&1&i
\\
1&-i&-1&i
\end{array}
\right).
\end{eqnarray}
\end{subequations}
These generate a $48$-element subgroup of $O(4)$ isomorphic to $\text{GL}(2,\mathbb{Z}_3)$ \cite{gap}.
With the PSG representation, we construct a fourth order Landau functional that is invariant under this group. 
\begin{equation}
\mathcal{L}
= | \partial \Psi |^2 +  r | \Psi |^2 + u | \Psi |^4
+ a_1 \mathcal{I}_1
+ a_2 \mathcal{I}_2 
\label{eq:case1-GL}
\end{equation}
with
\begin{subequations}
\label{eq:case1-inv}
\begin{eqnarray}
\mathcal{I}_1
&= &
-6\rho_1^2\rho_2^2+\rho_1^4\textup{cos}4\theta_1+\rho_2^4\textup{cos}4\theta_2,
\\
\mathcal{I}_2
&=&
\rho_1^3 \rho_2 \textup{sin}2\theta_1 [\textup{cos}(\theta_1+\theta_2)+\textup{sin}(\theta_1+\theta_2)] 
\nonumber\\
&+& 
\rho_1\rho_2^3 \textup{sin}2\theta_2 [\textup{cos}(\theta_1+\theta_2)-\textup{sin}(\theta_1+\theta_2)] 
\nonumber\\
&+&
\rho_1^2\rho_2^2 \textup{sin}[2(\theta_1-\theta_2)].
\end{eqnarray}
\end{subequations}
Here we have used the parametrization $\psi_n=\rho_n e^{i\theta_n}(n=1,2)$ with real-valued $\rho_n$ and $\theta_n$. We will be using the same parametrization for the other cases as well.

With the Landau functional, symmetry breaking patterns in VBS phases are investigated via a mean-field approach in the confined regime ($\Psi \ne 0$). We distinguish different phases by observing changes in the mean-field energy profile and the dimerization pattern. The dimerization pattern is identified by computing the dimer occupation number $d$ [Eq. (\ref{eq:dimer-occ})] at each kagome link; $d$ is calculated by using the dual mapping rule [Eq. (\ref{eq:dual-mapping})] and the soft mode expansion [Eq. (\ref{eq:vison-field-exp}) in this case] with mean-field solution $\Psi_{MF}$. In plotting the dimerization pattern, we use red ($d>\bar{d}$), black ($d<\bar{d}$), and dashed ($d=\bar{d}$) lines. Remember that $\bar{d}=1/4$ is the average dimer occupation. In our plotting scheme, the thickness of the red and black lines is proportional to the deviation from the average, $|d-\bar{d}|$. 

Figure \ref{fig:case1-phases} shows a mean-field phase diagram in the $a_1$-$a_2$ plane of the above Landau theory. The diagram contains three different phases: VBS$_{1A}$, VBS$_{1B}$, and VBS$_{1C}$. 
\begin{itemize}
\item \underline{\it Phase A}. The VBS$_{1A}$ phase is the 12-site VBS featured with the dimer structures of the diamonds and the stars [Fig. \ref{fig:VBSpatterns} (a)]. In this phase, the red dimers with $d>\bar{d}$ respect the hardcore dimer constraint (each site is occupied by only one red dimer) since it is dominated by a particular singlet product state. 
\item \underline{\it Phases B and C}. In contrast to phase $A$, VBS$_{1B}$  and VBS$_{1C}$ phases consist of the black and dashed bonds without any red dimers. The absence of the red dimers implies that the phases are superpositions of various dimer product states without a clear symmetry breaking pattern. Particularly, the VBS$_{1C}$ has uniform dimer density $d=\bar{d}$ despite having a nonzero vison condensation $\Psi_{MF}\ne0$.\cite{comment}
In this phase, symmetry breaking is anticipated to occur when fluctuation effects beyond the mean-field are incorporated. The VBS$_{1B}$ may be understood as a variation from the VBS$_{1A}$ with the star dimer structures being melted by dimer resonance. 
\end{itemize}

\subsection{Case 2: $\varphi=\pi$}

\begin{figure*}
 \includegraphics[width=0.45\linewidth]{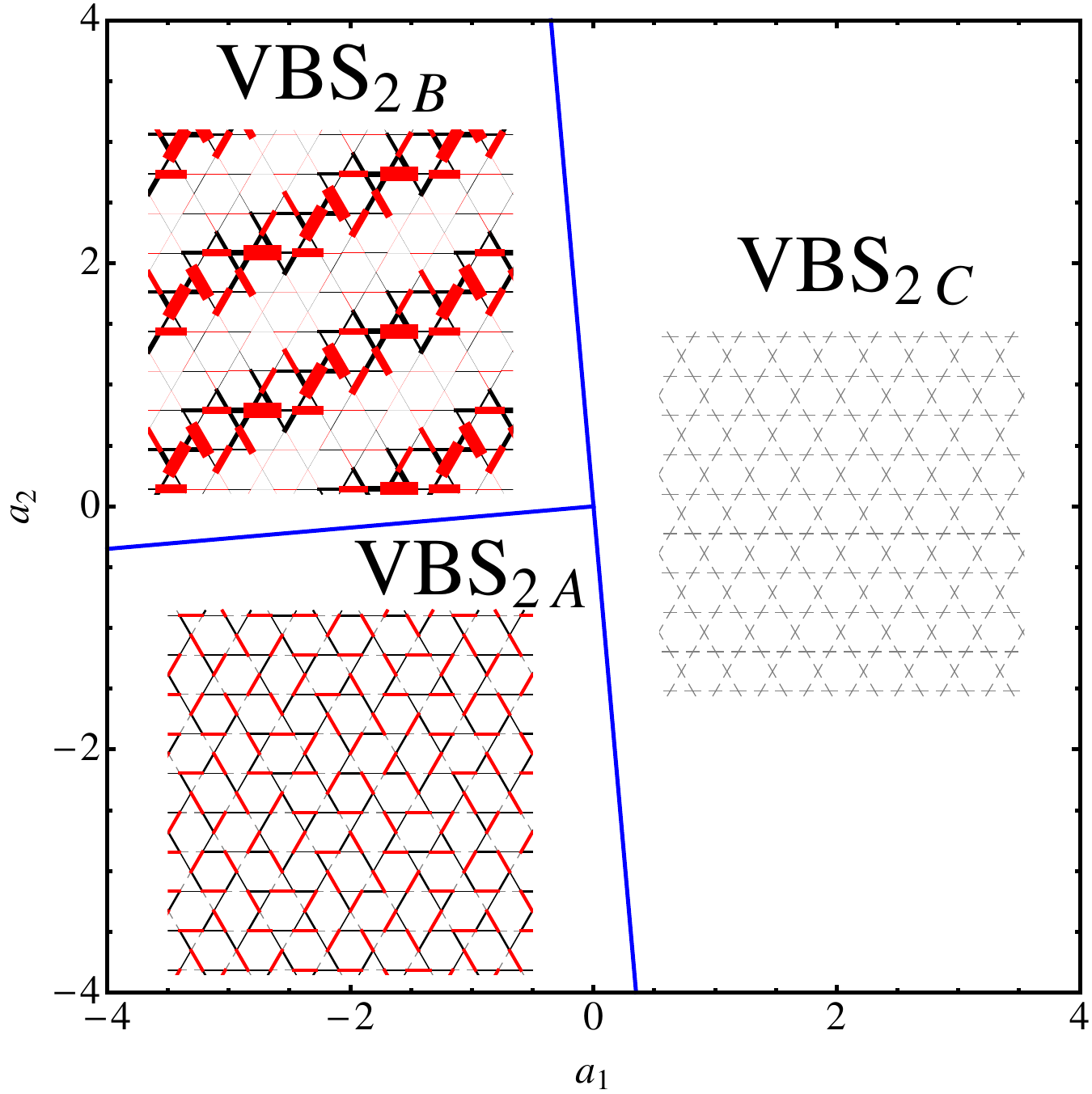}
 ~~~~~
 \includegraphics[width=0.45\linewidth]{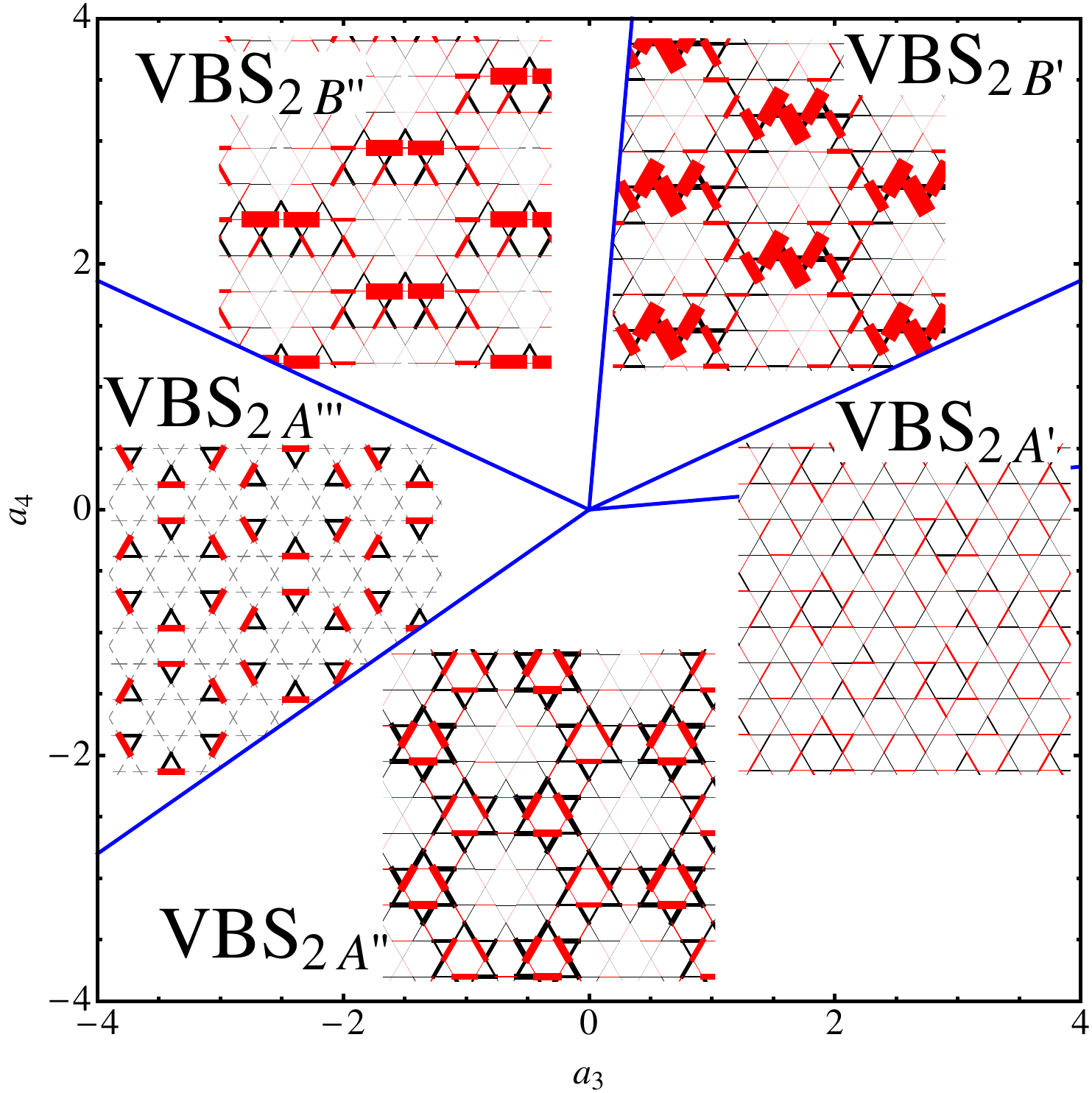}
 \caption{(Color online) Mean-field phase diagrams of the Landau theory [Eq. (\ref{eq:case2-GL})] for case 2.
 The phases VBS$_{2A}$, VBS$_{2A'}$, VBS$_{2A''}$, and VBS$_{2A'''}$ overall have the dimer structure of the 36-site VBS in Fig. \ref{fig:VBSpatterns} (b) with modifications in the dimer orientations in the hexagons and the overall dimer density. 
 The phases VBS$_{2B}$, VBS$_{2B'}$, and VBS$_{2B''}$ have their origin in the 6-site VBS in Fig. \ref{fig:VBSpatterns} (c) with different patterns of dimer density modulation. In each phase diagram, we set  $u=10$ and the other $a$-parameters to be zero ($a_{3,4}=0$ in the left and $a_{1,2}=0$ in the right).
 \label{fig:case2-phases}}
\end{figure*}

In case 2, there are eight soft modes at $\pm{\bf Q}_{1,2}$ with the symmetry group SG$_{1,2}$.
With the soft modes, vison field is expressed as
\begin{eqnarray}
\boldsymbol{\phi} ({\bf r}) 
&=& \left( \psi_1 {\bf v}^{(1)} + \psi_2 {\bf v}^{(2)}  \right) e^{i{\bf Q}_1\cdot{\bf r}} 
+ \textup{c.c.}
\nonumber\\
&+& \left( \psi_3 {\bf v}^{(1)} + \psi_4 {\bf v}^{(2)} \right) e^{i{\bf Q}_2\cdot{\bf r}} 
+ \textup{c.c.}
\end{eqnarray}
with the same eigenvectors ${\bf v}^{(1,2)}$ as in the case 1 and order parameters
$\Psi=(\psi_1,\psi_2,\psi_3,\psi_4,\psi_1^*,\psi_2^*,\psi_3^*,\psi_4^*)$.
The order parameters have the following matrix representation of the vison PSG.
\begin{subequations}
\label{eq:case2-PSG}
\begin{eqnarray}
\mathcal{T}_u
&=&
\left(
\begin{array}{cccc|cccc}
z&0&0&0&0&0&0&0
\\
0&z&0&0&0&0&0&0
\\
0&0&z^5&0&0&0&0&0
\\
0&0&0&z^5&0&0&0&0
\\
\hline
0&0&0&0&z^{-1}&0&0&0
\\
0&0&0&0&0&z^{-1}&0&0
\\
0&0&0&0&0&0&z^{-5}&0
\\
0&0&0&0&0&0&0&z^{-5}
\end{array}
\right),
\\
\mathcal{I}_y
&=&
\left(
\begin{array}{cccc|cccc}
0&0&0&0&1&0&0&0
\\
0&0&0&0&0&z^7&0&0
\\
0&0&0&0&0&0&1&0
\\
0&0&0&0&0&0&0&z^{11}
\\
\hline
1&0&0&0&0&0&0&0
\\
0&z^{-7}&0&0&0&0&0&0
\\
0&0&1&0&0&0&0&0
\\
0&0&0&z^{-11}&0&0&0&0
\end{array}
\right),
\\
\mathcal{R}_6
&=&
\frac{1}{2}
\left(
\begin{array}{cccc|cccc}
0&0&1&z^{11}&1&z^{11}&0&0
\\
0&0&z^2&z^7&z^8&z&0&0
\\
1&z^7&0&0&0&0&1&z^7
\\
z^{10}&z^{11}&0&0&0&0&z^4&z^5
\\
\hline
1&z&0&0&0&0&1&z
\\
z^4&z^{11}&0&0&0&0&z^{10}&z^5
\\
0&0&1&z^5&1&z^5&0&0
\\
0&0&z^8&z^7&z^2&z&0&0
\end{array}
\right),
\end{eqnarray}
\end{subequations}
where $z=e^{i\pi/6}$.
These matrices generate a $288$-element subgroup of $\text{O}(8)$ isomorphic to $\text{GL}(2,\mathbb{Z}_3) \times \text{D}(3)$ \cite{gap}.
With the PSG representation, the following fourth order Landau functional is obtained.
\begin{equation}
\mathcal{L}
= | \partial \Psi |^2 +  r | \Psi |^2 + u | \Psi |^4
+ \sum_{m=1}^4 a_m \mathcal{I}_m ,
\label{eq:case2-GL}
\end{equation}
where
\begin{subequations}
\begin{eqnarray}
\mathcal{I}_1
&=&\rho_1^2\rho_2^2+\rho_2^2\rho_3^2+\rho_3^2\rho_4^2+\rho_4^2\rho_1^2
\nonumber\\
&-&\rho_1^2\rho_3^2\left[1+\textup{cos}2(\theta_1+\theta_3)\right]
\nonumber\\
&-&\rho_2^2\rho_4^2\left[1+\textup{cos}2(\theta_2+\theta_4)\right] ,
\\
\mathcal{I}_2
&=&
\rho_1^2\rho_3^2+\rho_2^2\rho_4^2
\nonumber\\
&-&\rho_1\rho_2\rho_3\rho_4 \textup{cos}(\theta_1-\theta_2-\theta_3+\theta_4) 
\nonumber\\
&-&\rho_1\rho_2\rho_3\rho_4 \sqrt{3} \textup{sin}(\theta_1-\theta_2-\theta_3+\theta_4) ,
\\
\mathcal{I}_3
&=&
\mathcal{J}_1+\mathcal{J}_2-4(\sqrt{3}-1)\mathcal{J}_4 ,
\\
\mathcal{I}_4
&=&
-(\sqrt{3}+1)\mathcal{J}_2+\mathcal{J}_3+4\mathcal{J}_4 .
\end{eqnarray}
\label{eq:case2inv}
\end{subequations}
The explicit forms of $\mathcal{J}_{1,2,3,4}$ are relegated to Appendix \ref{app:inv} due to their lengthy expressions. 

Mean-field phase diagrams in this Landau theory are shown in Fig. \ref{fig:case2-phases}: in the $a_1$-$a_2$ plane (left) and $a_3$-$a_4$ plane (right). The phases appearing in the diagrams are classified into three groups: A, B, and C. The VBS phases are named after the groups the phases belong to; VBS$_{2A}$, VBS$_{2A'}$, VBS$_{2A''}$, and VBS$_{2A'''}$ are of type A. 
\begin{itemize}
\item \underline{\it Group A}. The phases in this group represent the 36-site VBS phases characterized by the dimer structures of the stars and hexagons [Fig. \ref{fig:VBSpatterns} (b)]. Among the phases, detailed dimer structures vary in aspects of the dimer orientations in the hexagons and the overall dimer density. To be specific, in VBS$_{2A}$ and VBS$_{2A'}$ two different dimer orientations are found in the hexagons whereas only one orientation is found in the hexagons of VBS$_{2A''}$. In terms of the dimer density, VBS$_{2A'}$ and VBS$_{2A''}$ have notable modulations in the dimer density across the red dimers while VBS$_{2A}$ has more or less uniform density in its red dimers. In the VBS$_{2A'''}$ phase, only the bridge dimers, the dimers connecting the stars and hexagons, are observed and the other star and hexagon dimer structures have melted away. It is interesting to note that VBS$_{2A'''}$ is similar to VBS$_{1B}$ in terms of the dimer melting.
\item \underline{\it Group B}. The three phases of this group have their origin in the 6-site VBS pattern composed of parallel and zigzag dimers [Fig. \ref{fig:VBSpatterns} (c)]. Here, the unit cell is enlarged by a dimer density modulation to 36 sites. The phases VBS$_{2B}$, VBS$_{2B'}$, and VBS$_{2B''}$ have different modulation patterns. In VBS$_{2B'}$ and VBS$_{2B''}$, the dimer density is concentrated on the parallel and zigzag dimers, respectively. In VBS$_{2B}$, the modulation occurs in both types of dimers.
\item \underline{\it Group C}. VBS$_{2C}$ has no clear dimerization pattern since the dimer density is $d=\bar{d}$ at every link. This is essentially the same phase as VBS$_{1C}$ at the mean-field level.
\end{itemize}

\subsection{Case 3: $0< \varphi \lesssim \frac{\pi}{3}$}

Now we consider case 3 that has four soft modes at $\pm{\bf Q}$ with the symmetry group SG$_{3,4}$.
In this case, the soft mode expansion of the vison field is exactly the same as in case 1 given by Eq. (\ref{eq:vison-field-exp}). The only difference between the two cases is the symmetry group, which is now SG$_{3,4}$ with the lowered rotation symmetry $R_3$. We find that the Landau functional with the symmetry group remains the same as that of case 1 up to sixth order. Therefore, the mean-field phases in case 3 are the same as those in case 1 to this order. This fact suggests that the 12-site VBS phase is robust against the triangle inequivalence.

\subsection{Case 4: $\frac{\pi}{2} \lesssim \varphi < \pi $ \label{subsec:case4}}

\begin{figure}
 \includegraphics[width=\linewidth]{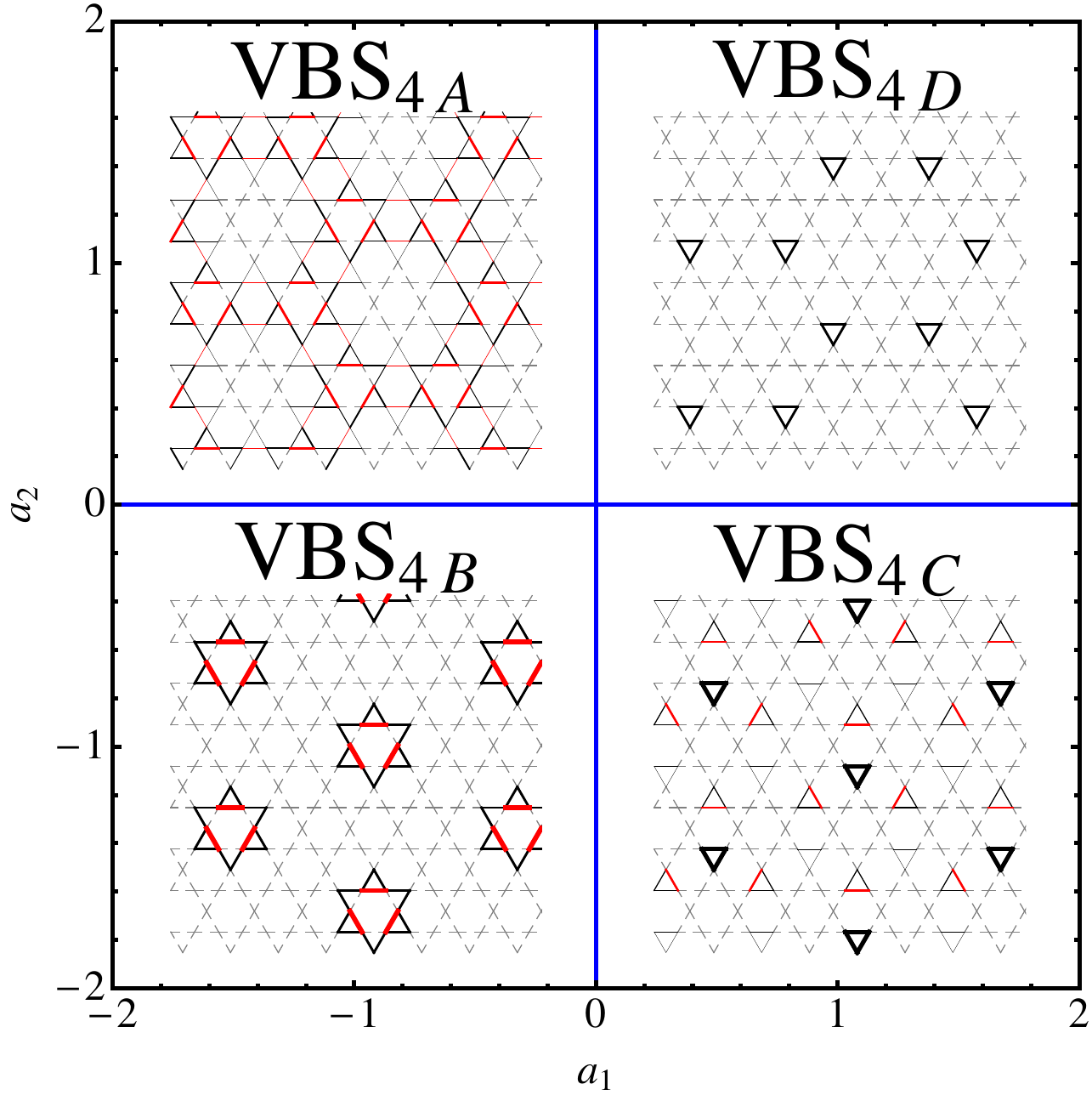}
 \caption{(Color online) Mean-field phase diagram of the Landau theory [Eq. (\ref{eq:case4-GL})] for case 4. In the four phases of the diagram, dimer melting effects, indicated by dashed lines ($d=\bar{d}$), are observed in common. The VBS$_{4A}$ and VBS$_{4B}$ are considered as variants of the 36-site VBS by the dimer melting. The same phase diagram is obtained in two different cases: (i)  $u=0,~v=20$, and (ii) $r=0,~v=10$.
 \label{fig:case4-phases}}
\end{figure}

In case 4, there are four soft modes at $\pm{\bf Q}_{1,2}$ under the symmetry group SG$_{3,4}$. Landau theory for this case is developed by the following soft mode expansion:
\begin{equation}
\boldsymbol{\phi} ({\bf r}) 
= \left(\psi_1 {\bf v}^{(1)} e^{i{\bf Q}_1\cdot{\bf r}} 
+ \psi_2 {\bf v}^{(2)} e^{i{\bf Q}_2\cdot{\bf r}} \right) 
+ \textup{c.c.}
\end{equation}
Here the eigenvectors, ${\bf v}^{(1,2)}$ take different forms from the previous cases:
\begin{equation}
{\bf v}^{(1)}=\frac{1}{\nu_1}
\left(
\begin{array}{c}
2-\sqrt{3}+i
\\
2
\end{array}
\right),
~~~~~
{\bf v}^{(2)}=\frac{1}{\nu_2}
\left(
\begin{array}{c}
2+\sqrt{3}+i
\\
2
\end{array}
\right),
\end{equation}
with the normalization factors, $\nu_1=2\sqrt{3-\sqrt{3}}$ and $\nu_2=2\sqrt{3+\sqrt{3}}$.
The order parameters, $\Psi=(\psi_1,\psi_2,\psi_1^*,\psi_2^*)$, have the following PSG representation.
 \begin{subequations}
 \begin{eqnarray}
\mathcal{T}_u
&=&
\left(
\begin{array}{cc|cc}
z&0&0&0
\\
0&z^5&0&0
\\
\hline
0&0&z^{-1}&0
\\
0&0&0&z^{-5}
\end{array}
\right) ,
\\
\mathcal{I}_y
&=&
\left(
\begin{array}{cc|cc}
0&0&z^7&0
\\
0&0&0&z^{11}
\\
\hline
z^{-7}&0&0&0
\\
0&z^{-11}&0&0
\end{array}
\right),
\\
\mathcal{R}_3
&=&
\frac{1}{\sqrt{2}}
\left(
\begin{array}{cc|cc}
z^{5/2}&0&0&z^4
\\
0&z^{-11/2}&z^2&0
\\
\hline
0&z^{-4}&z^{-5/2}&0
\\
z^{-2}&0&0&z^{11/2}
\end{array}
\right) ,
\end{eqnarray}
\end{subequations}
where $z=e^{i\pi/6}$.
The group generated is isomorphic to $(\text{C}_3 \times \text{SL}(2,\mathbb{Z}_3) ) \ltimes \text{C}_2$, \cite{gap} which is a finite subgroup of $\text{O}(4)$ with 144 elements.
The PSG representation leads to the following sixth order Landau functional:
\begin{equation}
\mathcal{L}
=
| \partial \Psi |^2 +  r | \Psi |^2 + u | \Psi |^4  + v | \Psi |^6
+ a_1 \mathcal{I}_1
+ a_2 \mathcal{I}_2 
\label{eq:case4-GL}
\end{equation}
with
\begin{subequations}
\begin{eqnarray}
\mathcal{I}_1
&=&
(\rho_1^2 - \rho_2^2) \rho_1^2 \rho_2^2 \textup{cos}2(\theta_1+\theta_2) ,
\\
\mathcal{I}_2
&=&
\rho_1^5 \rho_2 \textup{cos}(5\theta_1-\theta_2) + \rho_1 \rho_2^5 \textup{cos}(\theta_1-5\theta_2) .
\end{eqnarray}
\end{subequations}

Figure \ref{fig:case4-phases} shows four different mean-field phases in the $a_1$-$a_2$ plane. In all of the four phases, dimer melting effects, indicated by dashed lines ($d=\bar{d}$), are observed. 
\begin{itemize}
\item \underline{\it Phases A and B}. VBS$_{4A}$ and VBS$_{4B}$ are considered as variants of the 36-site VBS. VBS$_{4A}$ possesses hexagon and bridge dimer structures although the stars are all erased out. In VBS$_{4B}$, most of the 36-site VBS structures are wiped out leaving a triangular arrangement of the hexagon structures.
\item \underline{\it Phases C and D}. VBS$_{4C}$ and VBS$_{4D}$ have similar structures to those of VBS$_{2A'''}$ and VBS$_{1B}$. Red dimers are found scattered in the VBS$_{4C}$. Interestingly, the red dimers do not fit to any characteristic dimer structure observed in the 12-site, 36-site, and 6-site VBS phases.
\end{itemize}

\section{Discussions\label{sec:discussion}}

In this paper, we studied the $\mathbb{Z}_2$ gauge theory represented by the Hamiltonian $H_{ZGT}$ [Eq. (\ref{eq:Hamiltonian})] as an effective theory for low energy spin-singlet fluctuations in kagome lattice antiferromagnets. Our study was focused on valence bond solid phases in the intermediate regime sandwiched between the two exactly solvable points in the phase diagram of Fig. \ref{fig:schematic-PD}. It is found that the 12-site VBS phase with the diamond pattern is the only VBS phase that can be realized on the $\varphi=0$ line. This is supported by three different approaches: the ground state manifold of the Hamiltonian when $h=0$, the wave vectors of the lowest two-vison excitations, and the dimerization pattern from the Landau theory (VBS$_{1A}$). According to recent DMRG simulations,\cite{Yan_2011} the 12-site VBS is considered the long-range ordered counterpart of the $\mathbb{Z}_2$ spin liquid ground state of the spin-1/2 nearest-neighbor Heisenberg model.

On the other hand, two different types of VBS phases are found on the $\varphi=\pi$ line of the phase diagram. One of the phases is the 36-site VBS that was proposed as a ground state candidate of the Heisenberg model based on various studies, including the quantum dimer model\cite{Nikolic_2003} and the series expansion.\cite{Singh_2007} We have identified three 36-site VBS patterns, VBS$_{2A}$, VBS$_{2A'}$, and VBS$_{2A''}$. The remaining phase is the 6-site VBS. In our Landau theory, the phase appears with an enlarged 36-site unit cell by dimer density modulations as in the phases VBS$_{2B}$, VBS$_{2B'}$, and VBS$_{2B''}$. It is interesting to note that the 6-site VBS does not have at all the hexagon and doubled-diamond dimer structures favored by the dimer interaction when $K<0$ [compare Fig. \ref{fig:VBSpatterns} (c) and Fig. \ref{fig:dimer-motion-int} (c)]. It contrasts with the 36-site VBS that contains those two dimer structures.

\begin{table}
\begin{ruledtabular}
\begin{tabular}{lc}
Our theory & Ref. [\onlinecite{Huh_2011}]
\\
\hline
VBS$_{1A}$ & VBS 1$_{F}$
\\
VBS$_{1B}$ & VBS 2$_{F}$
\\
VBS$_{2A,2A'}$ & VBS 1$_{A}$
\\
VBS$_{2A'''}$ & VBS 2$_{A}$
\\
VBS$_{2B'}$ & VBS 4$_{A}$
\\
VBS$_{2B''}$ & VBS 3$_{A}$  
\end{tabular}
\end{ruledtabular}
\caption{Correspondence between the results of this paper and Ref. [\onlinecite{Huh_2011}]
\label{tab:VBS-comparison}}
\end{table}

The above results are remarkably consistent with a previous work.\cite{Huh_2011} In a different form of the $\mathbb{Z}_2$ gauge theory, they proposed 12-site VBS and 36-site VBS states as promising VBS phases in confinement transitions from the $\mathbb{Z}_2$ spin liquid phase. While the representations of the visons differ in the two theories, the group under which the visons transform are isomorphic to one another in each case. The $\varphi=0$ ($\varphi=\pi$) corresponds to the ferromagnetic (antiferromagnetic) next nearest neighbor interaction considered in Ref. [\onlinecite{Huh_2011}]. Also, as summarized in Table \ref{tab:VBS-comparison}, there is a surprising correspondence between the VBS patterns. Interestingly, the 6-site VBS patterns, VBS$_{2B'}$ and VBS$_{2B''}$, in our theory also appear in the other theory.

The appearance of the 6-site VBS state in the Landau theory approach as well as in the ground state manifold of $H_{ZGT}$ with $K<0$ and $h=0$ suggests that it may be a competing low energy state considered in the kagome antiferromagnets. In fact, it was hinted in the DMRG simulations of the Heisenberg model. On the families of cylinders, YC$4m+2$ and YC$(4m+1)-2$, the ground state of the Heisenberg model displays symmetry-breaking valence bond modulation patterns; strong valence bonds formed along zigzag lines and straight lines. The modulation patterns are understood as a resonating state of the 6-site VBS dimer configurations with the zigzag and straight lines as transition graphs [see Fig. \ref{fig:VBSpatterns} (c)].

Beyond the ideal kagome lattice geometry, we also investigated effects of the triangle inequivalence on VBS order formation. Upon introducing inequivalent triangles via the dimer potential energy term $A$, the 12-site VBS and 36-site VBS respond in different ways. Around the line $\varphi=0$ of Fig. \ref{fig:schematic-PD}, the 12-site VBS remains stable to anisotropy as suggested by the unaffected (i) ground state manifold of the Hamiltonian with $h=0$, (ii) vison soft modes, and (iii) Landau theory with the lowered rotation symmetry (the case 3).

Around the $\varphi=\pi$ line, the inequivalent triangles have dimer melting effects on the 36-site VBS phase as represented by the dimerization patterns in VBS$_{4A}$ and VBS$_{4B}$. In both patterns, the star dimer structures are all melted whereas the hexagon structures remain at least partly. This trend can be understood by noting that the hexagon structures carry empty triangles
energetically favored by the dimer potential $A$ while the star structures do not have any empty triangles. Hence, the hexagons are more stable. At the stars, dimer resonances, which lead to dimer melting, are enhanced by the inequivalent triangle geometry. Interestingly, this is in contrast to the stability of the 12-site VBS (triangular array of the stars) around the $\varphi=0$ line. The different behaviors at $\varphi\simeq0$ and $\varphi\simeq\pi$ are attributed to their different structures of low energy space of the Hamiltonian. We also note that no analogue of the 6-site VBS is found in the inequivalent triangle geometry.

We have also attempted to investigate effects of the triangle anisotropy by extending the previous $\mathbb{Z}_2$ gauge theory in Ref. [\onlinecite{Huh_2011}]. As discussed earlier, both versions of $\mathbb{Z}_2$ gauge theory capture the 12-site and 36-site VBS phases on the ideal kagome lattice. As we tune anisotropy, both theories predict regions of 12-site VBS, incommensurate VBS, and 36-site VBS phases. They both show similar symmetry breaking patterns by the triangle anisotropy on the 36-site VBS phase. However, the 12-site VBS is destroyed immediately in favor of the incommensurate phase in the conventional theory, while it survives up to a finite critical anisotropy in our theory. The latter difference is ascribed to difference in the way the triangle anisotropy is incorporated in each theory. Details on the conventional $\mathbb{Z}_2$ gauge theory with triangle anisotropy can be found in Appendix~\ref{app:comparison}.

\begin{table}
\begin{ruledtabular}
\begin{tabular}{lcccr}
 VBS & size of unit cell & $R_6$ & $R_3$ & $I_y$
 \\
 \hline
 $1A$ & 12 & \checkmark & \checkmark &
 \\
 $1B$ & 12 & \checkmark & \checkmark & \checkmark
 \\
 $1C$ & 3 & \checkmark & \checkmark & \checkmark
 \\
 \hline
 $2A, 2A', 2A'', 2A'''$ & 36 & & \checkmark &
 \\
 $2B, 2B', 2B''$ & 36 & & &
 \\
 $2C$ & 3 & \checkmark & \checkmark & \checkmark
 \\
 \hline
 $4A, 4B, 4C, 4D$ & 36 & & \checkmark & \checkmark
\end{tabular}
\end{ruledtabular}
\caption{Remaining symmetries in the VBS phases.\label{tab:remaining-symm}}
\end{table}

Now we discuss remaining symmetries in the VBS phases found in this work. For each phase, size of unit cell and remaining point group symmetries are summarized in Table \ref{tab:remaining-symm}. In the 12-site VBS phase ($1A$), the $R_6$ rotation symmetries with respect to the centers of the dimer stars are preserved while the $I_y$ inversions are completely broken. In comparison with the 12-site VBS, the 36-site VBS phases ($2A$, $2A'$, $2A''$) have less symmetries due to more complicated and larger unit cell: only the $R_3$ rotations  with respect to the centers of the dimer stars among the point group symmetries. In other 36-site VBS phases ($2B$, $2B'$, $2B''$), when considered as dimer-modulated 6-site VBS states, all the point group symmetries are completely broken. In contrast to the above phases, the phases $1C$ and $2C$ may seem to have no broken symmetries with uniform dimer density. However, these uniform dimer density phases are results obtained at the mean-field level. We expect symmetry breaking pattern would be emergent when fluctuation effects are incorporated in those phases. In the anisotropic kagome lattice case, the VBS phases, $4A$, $4B$, $4C$, $4D$, have the $I_y$ inversions as well as the $R_3$ rotations. It is interesting to note that the VBS phases on the anisotropic kagome lattice preserve more symmetries compared to the 36-site VBS phases found on the isotropic kagome lattice.

In this work, we considered a simple and generic quantum dimer model that could describe low energy singlet physics of kagome antiferromagnets. In regard to antiferromagnets on the anisotropic kagome lattice, it would be an interesting problem to investigate which microscopic spin models can be mapped onto our effective quantum dimer model. In the present situation without any numerical studies on anisotropic kagome lattice models, Mila's effective model construction\cite{Mila_1998} on the $J$-$J'$ anisotropic kagome Heisenberg model can be helpful for understanding the connection between the quantum dimer model and $J$-$J'$ Heisenberg model. It would also be interesting to compare Mila's effective model with our quantum dimer model on the anisotropic kagome lattice. We leave the investigation in this direction as a future study.

\acknowledgments

We thank Robert Schaffer for helpful discussions and collaborations on a related study. This work was supported by the NSERC of Canada, the Canadian Institute for Advanced Research, and the Center for Quantum Materials at the University of Toronto. Some parts of this work (YBK) were performed at the KITP, supported by NSF PHY11-25915.

\appendix

\section{Exactly solvable points at $h=0$\label{app:zeroh}}

\begin{figure}
\centering
 \includegraphics[width=0.7\linewidth]{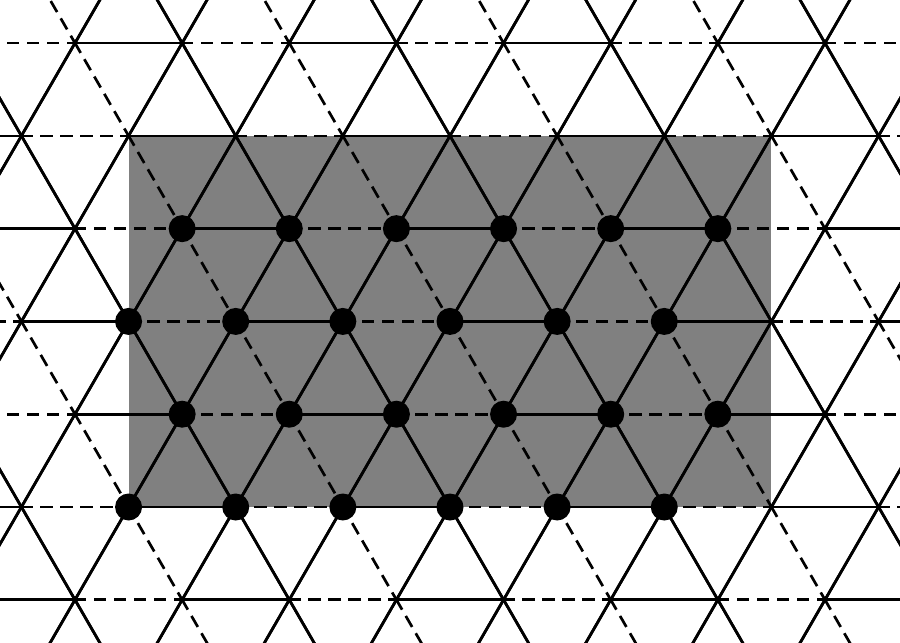}
 \includegraphics[width=\linewidth,angle=270]{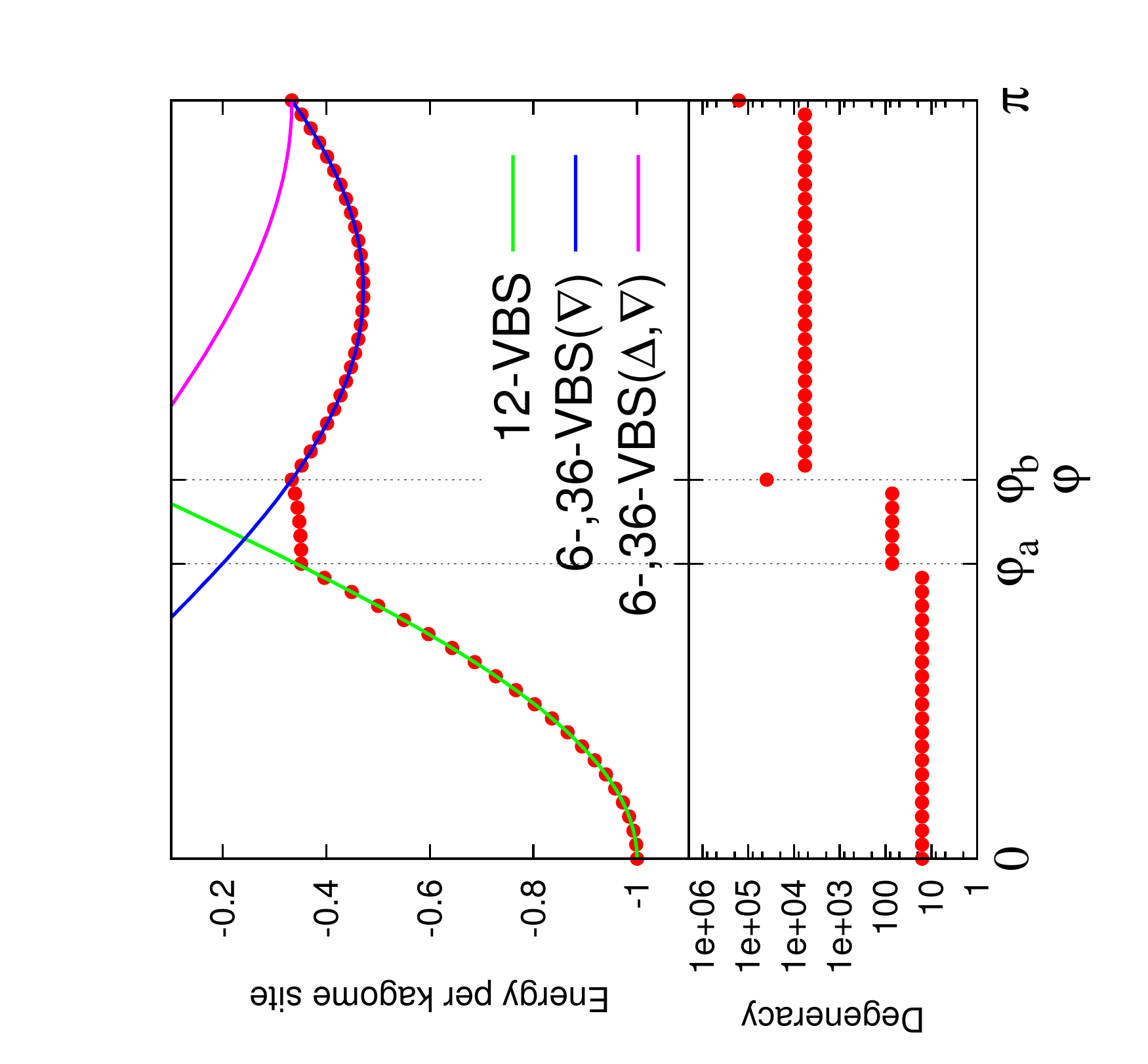}
 \caption{(Color online) Plots of the ground state energy of the Hamiltonian at $h=0$ and degeneracy as functions of the angle parameter $\varphi$. The results are obtained with a full search in the 24-site cluster (at the top) with periodic boundary conditions. For explanations of the plots, see Appendix \ref{app:zeroh}.
\label{fig:zerohpoints}}
\end{figure}

Here we discuss in more detail the $h=0~(\vartheta=\pi/2)$ exactly solvable points of the phase diagram in Fig. \ref{fig:schematic-PD}. At these points, the dual Ising model $H_{DIM}$ is readily diagonalized as follows.
\begin{equation}
 K \sum_{\langle\langle\langle \alpha, \gamma \rangle\rangle\rangle}  \lambda_{\alpha\beta} \lambda_{\beta\gamma} \tau_{\alpha}^z \tau_{\gamma}^z + A \sum_{\langle \alpha, \beta \rangle} \lambda_{\alpha\beta} \tau_{\alpha}^z \tau_{\beta}^z 
\end{equation}
with $K=\textup{cos}\varphi$ and $A=\textup{sin}\varphi$.
To explore the ground state manifold of the Hamiltonian, we put the system on a finite size cluster of the dual triangular lattice with a periodic boundary condition.

We design the cluster to capture the 12-site and 36-site VBS states in Fig. \ref{fig:VBSpatterns}.
It is easy to see that the 12-site VBS state is the exact ground state at $A=0$ by noting that the state consists of the star and diamond dimer configurations with the minimum energy of the $K$ potential [see Fig. \ref{fig:VBSpatterns} (a) and Table \ref{tab:trg}]. In the dual Ising model description, $\tau^z$-spin configurations representing the 12-site VBS have a 8-site unit cell on the triangular lattice, which constrains the cluster to have a multiple of 8 sites. 
On the other hand, possibility of the 36-site VBS state (strictly speaking, the size of unit cell) as the ground state was hinted in the soft-spin approximation approach (Sec. \ref{sec:soft}). VBS states with a 36-site unit cell require the cluster to have a multiple of 12 sites.
Combining the above two conditions, we find that the cluster size should be $24n$ with $n=1,2,\cdots$.
As a cluster satisfying this condition, we consider the 24-site cluster in Fig. \ref{fig:zerohpoints}, which corresponds to a 72-site cluster on the kagome lattice.
By performing a full search on the cluster, we obtain the ground state energy and degeneracy as shown in the figure (red dots as functions of the angle parameter $\varphi$). In the parameter range $0\leq \varphi \leq \pi$, we find three different regions separated by two points, $\varphi_a=1.222$ and $\varphi_b=1.571$.

First, in the region $0\leq \varphi < \varphi_a$, the ground state manifold only contains dimer configurations of the 12-site VBS phase. In this case, the ground state energy per kagome site is $-K=-\textup{cos}\varphi$ (green line in the figure). One can easily check the energy by acting the QDM Hamiltonian $H_{QDM}$ with $ h=0$ on the dimer product state representing the 12-site VBS [Fig. \ref{fig:VBSpatterns} (a)].

When $\varphi_b < \varphi \leq \pi $, the ground state manifold consists of various VBS phases including the 36-site VBS and 6-site VBS. Depending on the presence/absence of the dimer potential energy ($A$) in the Hamiltonian, empty triangles in the VBS states are arranged in different ways. When $\varphi_b < \varphi < \pi $, the empty triangles are all down-triangles in both 36-site and 6-site VBSs. The energy of these states, $(K-A)/3=(\textup{cos}\varphi-\textup{sin}\varphi)/3$, is plotted in blue in the figure. However, at the point $\varphi = \pi$, the up- and down-triangles have no distinction due to the absence of dimer potential. This fact leads to more choices on the distribution of the empty triangles in the VBS, which is reflected in the increased ground state degeneracy at that point (see the lower plot in Fig. \ref{fig:zerohpoints}). The pink line represents the energy of the 36-site VBS and modified 6-site VBS states with the same number of up- and down-empty-triangles: $K/3=\textup{cos}\varphi/3$. 
The 36-site VBS is shown in Fig. \ref{fig:VBSpatterns} (b). The modified 6-site VBS is obtained by arranging the parallel dimers of the 6-site VBS [Fig. \ref{fig:VBSpatterns} (c)] in such a way that lines of the parallel dimers have alternating dimer configurations. Note that the empty-triangles are distributed along the parallel dimer lines in the 6-site VBS.

In the intermediate region $\varphi_a < \varphi < \varphi_b$, the dimer potential energy dominates the interaction energy, and none of the 12-site, 36-site, and 6-site VBS states are found. This region corresponds to the parameter region $\frac{\pi}{3} \lesssim |\varphi| \lesssim \frac{\pi}{2}$ of $H_{soft}$ where vison soft modes occur at incommensurate momentum positions (Sec. \ref{sec:soft}).

\section{Vison PSG\label{app:visonPSG}}

In this appendix, we provide explicit expressions for the generators of the vison PSG in Fig. \ref{fig:visonPSG}. First, physical symmetry part of each generator on the dual triangular lattice is expressed as follows. 
\begin{subequations}
\begin{eqnarray}
T_u &:& (m,n)_s \rightarrow (m+1,n)_s,
\\
\nonumber\\
I_y 
& : &
 \left\{
 \begin{array}{l}
 (m,n)_1 \rightarrow (-m,n)_1
 \\
 (m,n)_2 \rightarrow (-m-1,n)_2
 \end{array}
 \right. ,
\\
\nonumber\\
\nonumber\\
R_6 
& : &
 \left\{
 \begin{array}{lcl}
 (m,n)_1 \rightarrow (\frac{m-3n}{2},\frac{m+n}{2})_1 && (m+n:\textup{even})
 \\
 (m,n)_2 \rightarrow (\frac{m-3n-2}{2},\frac{m+n}{2})_2 && (m+n:\textup{even})
 \\
 &&
 \\
 (m,n)_1 \rightarrow (\frac{m-3n-1}{2},\frac{m+n-1}{2})_2 && (m+n:\textup{odd})
 \\
 (m,n)_2 \rightarrow (\frac{m-3n-1}{2},\frac{m+n+1}{2})_1 && (m+n:\textup{odd})
 \end{array}
 \right. ,
\nonumber\\
\\
\nonumber\\
 R_3 
 & : &
 \left\{
 \begin{array}{lcl}
 (m,n)_1 \rightarrow (\frac{-m-3n}{2},\frac{m-n}{2})_1 && (m+n:\textup{even})
 \\
 (m,n)_2 \rightarrow (\frac{-m-3n-2}{2},\frac{m-n}{2})_1 && (m+n:\textup{even})
 \\
 &&
 \\
 (m,n)_1 \rightarrow (\frac{-m-3n-1}{2},\frac{m-n-1}{2})_2 && (m+n:\textup{odd})
 \\
 (m,n)_2 \rightarrow (\frac{-m-3n-3}{2},\frac{m-n-1}{2})_2 && (m+n:\textup{odd})
 \end{array}
 \right. .
 \nonumber\\
\end{eqnarray}
\end{subequations}
Here, the notation $(m,n)_s$ represents the sublattice site $s~(=1,2)$ in the unit cell located at $m{\bf u}+n{\bf v}~(m,n\in\mathbb{Z})$.
For the above symmetry operations, the accompanied gauge transformations can be chosen as follows.
\begin{subequations}
\begin{eqnarray}
G_{T_u} (m,n)_s & = & 1,
\\
G_{I_y} (m,n)_s
& = &
\left\{
\begin{array}{lcl}
(-1)^n && (s=1)
\\
(-1)^{n+1} && (s=2)
\end{array}
\right. ,
\\
G_{R_6} (m,n)_s
& = & 
\left\{
\begin{array}{lcl}
(-1)^m && (s=1)
\\
(-1)^n && (s=2)
\end{array}
\right. ,
\\
G_{R_3} (m,n)_s 
& = &
\left\{
\begin{array}{cc}
 -1 & 
 \left(
\begin{array}{c}
s=1;~\textup{mod}(m+n,4)=0,1
\\
s=2;~\textup{mod}(m+n,4)=2,3
\end{array}
\right)
 \\
 1 & (\textup{otherwise})
\end{array}
\right.
 .
 \nonumber\\
\end{eqnarray}
\end{subequations}

\section{Invariants $\mathcal{I}_{3,4}$ in the case 2\label{app:inv}}

The invariants $\mathcal{I}_{3,4}$ in Eq. (\ref{eq:case2inv} c,d) are composed of four terms $\mathcal{J}_{1,2,3,4}$:
\begin{widetext}
\begin{subequations}
\begin{eqnarray}
\mathcal{J}_1
&=&
\rho_1^2 \rho_3 \rho_4 
\left[
\textup{cos} (2\theta_1 + \theta_3 + \theta_4)
+
(-2+\sqrt{3})
\textup{sin} (2\theta_1 + \theta_3 + \theta_4)
\right]
\nonumber\\
&+&
\rho_2^2 \rho_3 \rho_4 
\left[
-\textup{cos} (2\theta_2 + \theta_3 + \theta_4)
+
(-2+\sqrt{3})
\textup{sin} (2\theta_2 + \theta_3 + \theta_4)
\right]
\nonumber\\
&+&
\rho_1 \rho_2 \rho_3^2 
\left[
(-2+\sqrt{3})
\textup{cos} (\theta_1 + \theta_2+2\theta_3)
+
\textup{sin} (\theta_1 + \theta_2+2\theta_3)
\right]
\nonumber\\
&+&
\rho_1 \rho_2 \rho_4^2 
\left[
(2-\sqrt{3})
\textup{cos} (\theta_1 + \theta_2+2\theta_4)
+
\textup{sin} (\theta_1 + \theta_2+2\theta_4)
\right],
\\
\nonumber\\
\mathcal{J}_2
&=&
(\rho_1^3 \rho_2  - \rho_1 \rho_2^3) 
\left[
(2-\sqrt{3})
\textup{cos} (\theta_1 - \theta_2)
+
\textup{sin} (\theta_1 - \theta_2)
\right]
\nonumber\\
&-&
(\rho_3^3 \rho_4  - \rho_3 \rho_4^3) 
\left[
\textup{cos} (\theta_3 - \theta_4)
+
(2-\sqrt{3})
\textup{sin} (\theta_3 - \theta_4)
\right],
\\
\nonumber\\
\mathcal{J}_3
&=&
\rho_1^2 \rho_2^2
\left[
-\sqrt{3} \textup{cos}2(\theta_1-\theta_2) + \textup{sin}2(\theta_1-\theta_2)
\right]
\nonumber\\
&+&
\rho_3^2 \rho_4^2
\left[
\sqrt{3} \textup{cos}2(\theta_3-\theta_4) + \textup{sin}2(\theta_3-\theta_4)
\right]
\nonumber\\
&+&
(\rho_1 \rho_2 \rho_3^2-\rho_1 \rho_2 \rho_4^2)
\left[
(-1+\sqrt{3}) \textup{cos}(\theta_1-\theta_2) + (1+\sqrt{3}) \textup{sin}(\theta_1-\theta_2)
\right]
\nonumber\\
&+&
( -\rho_1^2 \rho_3 \rho_4+\rho_2^2 \rho_3 \rho_4 )
\left[
(1+\sqrt{3}) \textup{cos}(\theta_3-\theta_4) + (-1+\sqrt{3}) \textup{sin}(\theta_3-\theta_4)
\right],
\\
\nonumber\\
\mathcal{J}_4
&=&
\rho_1 \rho_2 \rho_3 \rho_4 \textup{sin} (\theta_1-\theta_2+\theta_3-\theta_4).
\end{eqnarray}
\end{subequations}
\end{widetext}

\section{Comparison with the conventional gauge theory\label{app:comparison}}

\begin{figure}[h]
\includegraphics[width=30mm]{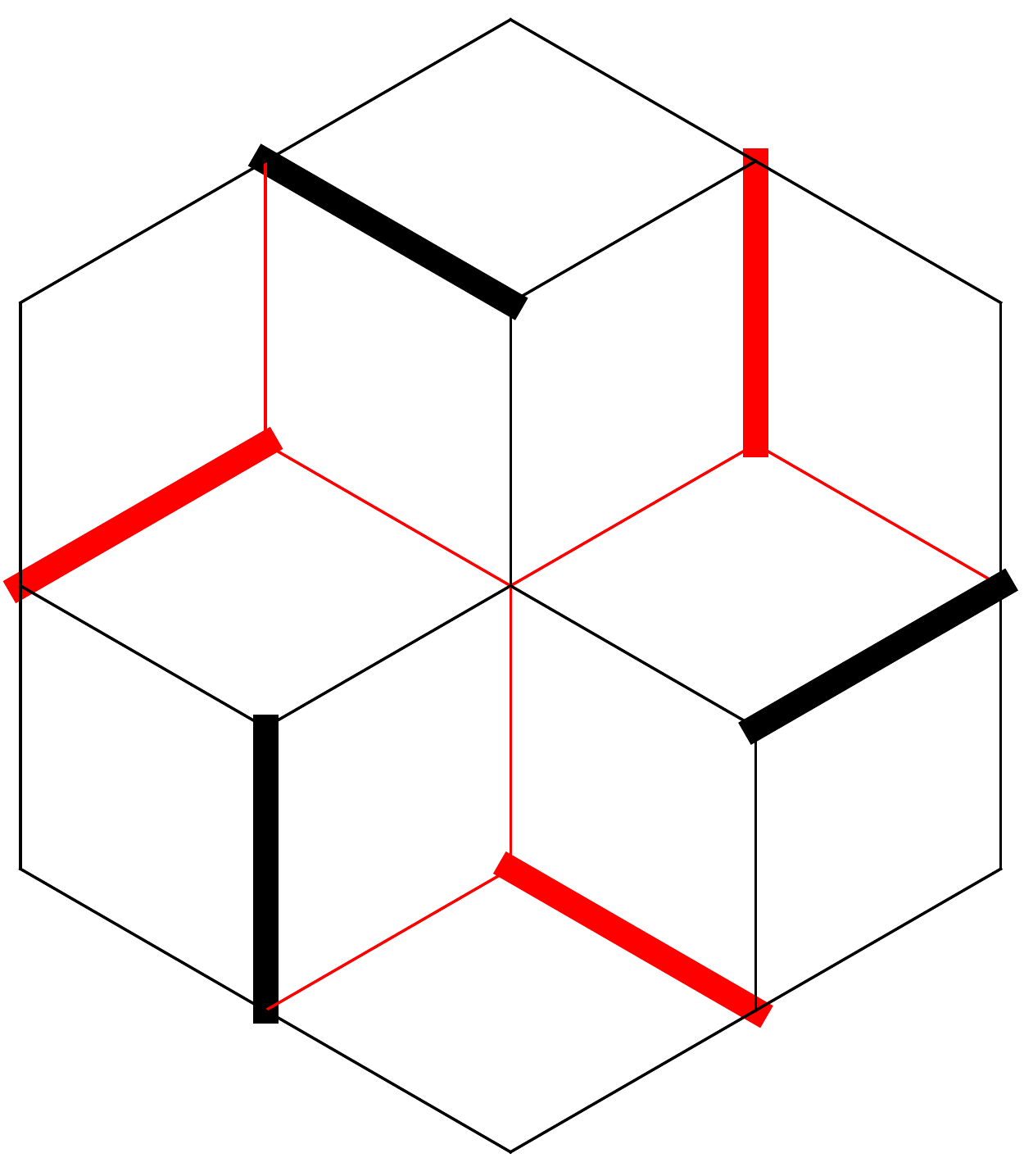}
\quad\quad
\includegraphics[width=30mm]{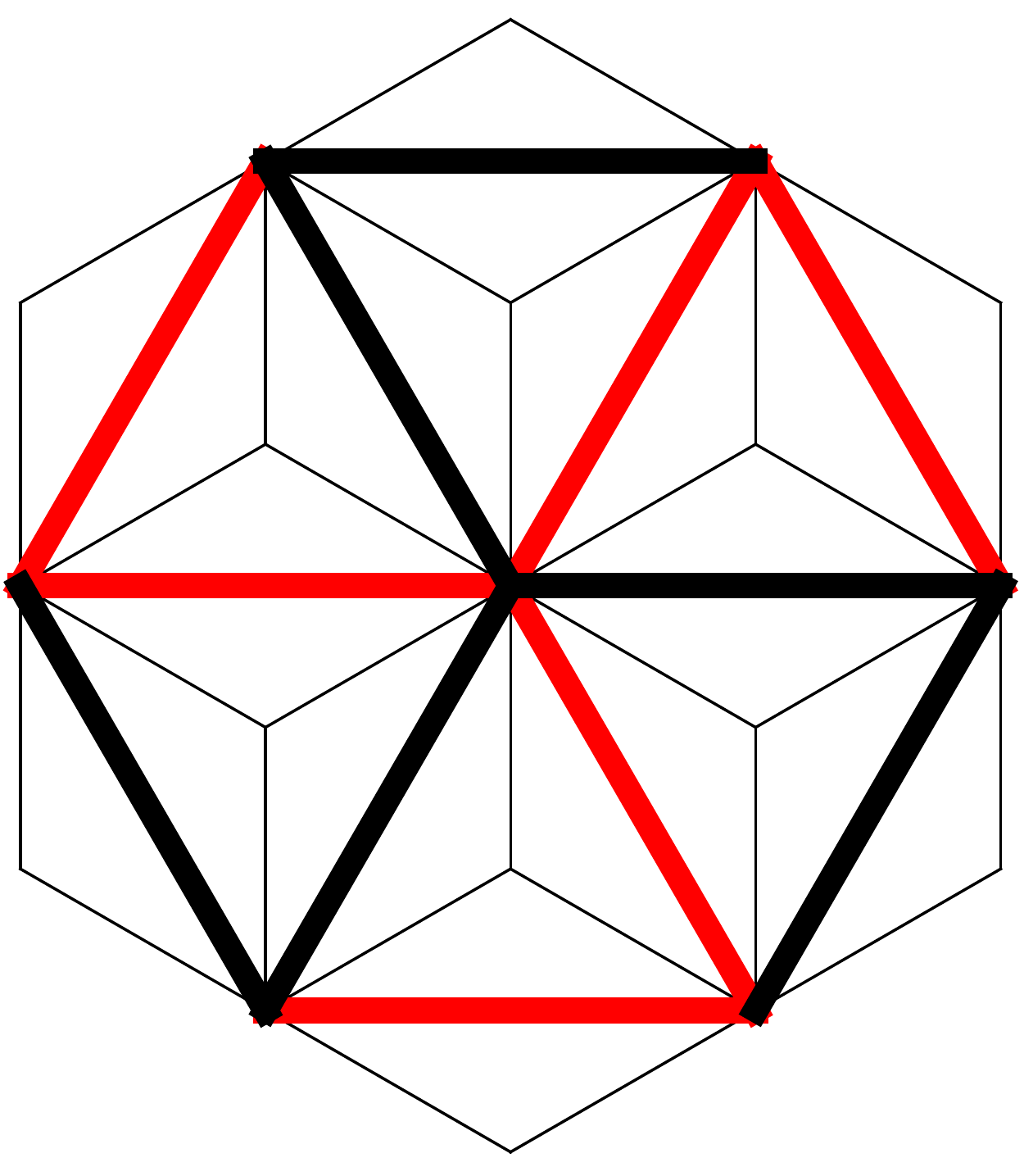}
\caption{(Color online) (left) Unit cell of a fully frustrated dice lattice with nearest neighbor interactions. It has 12 sites in the unit cell and the black (red) lines denote coupling strength $J_1$ ($J_2$). Thick bonds have negative sign relative to the thin bonds. (right) Next nearest neighbor interaction ($t_2$) that is only allowed when the $6$-fold rotation symmetry is broken down to $3$-fold. Thick red bonds have negative sign relative to the thick black bonds. }
\label{fig:dice}
\end{figure}

Here we compare the results of the anisotropic kagome lattice with traditional $Z_2$ gauge theory used in Ref.~[\onlinecite{Huh_2011}]. In this, the visons live on a fully frustrated dice lattice, corresponding to centers of the triangles and hexagons of the kagome lattice. (Fig.~\ref{fig:dice}) As noted in Ref.~[\onlinecite{Wan_2013}], the visons at the centers of triangles are higher in energy than the ones in the hexagons and the lowest energy description can be done using the visons on a triangle lattice. The low energy results of both theories should be equivalent and is confirmed below in our case.  

The Hamiltonian of the visons can be written as
\begin{equation}
\mathcal{H} = -\sum_{i,j} J_{ij}\phi_i \phi_j \;,
\end{equation}
where the couplings $J_{ij}$'s are chosen such that the products around a plaquette are negative. With anisotropy, we have two independent nearest neighbor vison interactions, denoted $J_1$ and $J_2$. These correspond to interaction between visons living in the center of hexagon and up/down triangles. When $J_1 = J_2$ the vison bands are flat in momentum space, and in Ref.~[\onlinecite{Huh_2011}] ferromagnetic or antiferromagnetic next nearest neighbor interactions connecting two $3$-coordinated sites ($t$) were added to give dispersion to the flat vison bands. $J_1 \neq J_2$ immediately gives dispersion to the visons and is smoothly connected to the antiferromagnetic $t$-interaction considered in Ref.~[\onlinecite{Huh_2011}]. The states resulting from ferromagnetic next nearest neighbor interactions were immediately destroyed (which is different from the results of the main text) and the soft modes were found to be incommensurate to the lattice. This corresponds to states found in the main text at $\pi/3 \lesssim \varphi \lesssim \pi/2$. Here we only consider the $t \le 0$ case.

The PSG found for the vison modes can be represented by these matrices. 
\begin{align}
T_u = \frac{1}{\sqrt{3}} \left( \begin{array}{cccc} 
	e^{i\pi/6} & \sqrt{2} e^{i 5 \pi/6} & 0 & 0 \\
	-i \sqrt{2}  & -e^{i\pi/6} & 0 & 0 \\
	0 & 0 & e^{-i\pi/6} & \sqrt{2} e^{-i 5 \pi/6} \\
	0 & 0 & i \sqrt{2} & -e^{-i\pi/6}
	\end{array}	\right)	\nonumber \\
I_x = \left( \begin{array}{cccc}
	0 & 0 & -1 & 0 \\
	0 & 0 & 0 & -1 \\
	-1 & 0 & 0 & 0 \\
	0 & -1 & 0 & 0 
	\end{array} \right) ~
R_3 = \left( \begin{array}{cccc}
	-e^{i \pi/3} & 0 & 0 & 0 \\
	0  & 1 & 0 & 0 \\
	0 & 0 & e^{i 2 \pi/3} & 0 \\
	0 & 0 & 0 & 1
	\end{array} \right) 
\end{align}
These matrices generate a group isomorphic to $(C_3 \times SL(2,\mathbb{Z}_3) ) \ltimes C_2$, \cite{gap} which is a finite subgroup of $O(4)$ with 144 elements. Note that this is identical to the group structure found in Sec.~\ref{subsec:case4}.
To sixth order, the effective Lagrangian that is invariant under this group is 
\begin{align}
\mathcal{L}
&= 
| \partial \Psi |^2 +  r | \Psi |^2 + u | \Psi |^4  + v | \Psi |^6 \nonumber \\
& + a \left[  \rho_1^6 \cos{6 \theta_1} -   \rho_2^6 \cos{6 \theta_2} - 5\sqrt{2}  \rho_1^3 \rho_2^3 \cos{3(\theta_1+\theta_2)} \right] \nonumber \\
& + b \left[ \rho_1^6 - 9 \rho_1^4 \rho_2^2 + 9 \rho_1^2 \rho_2^4 - \rho_2^6 + 4\sqrt{2} \rho_1^3 \rho_2^3 \cos{3(\theta_1-\theta_2)}  \right] \;.
\label{eq:GL-action}
\end{align} 

\begin{figure}[h]
\includegraphics[width=80mm]{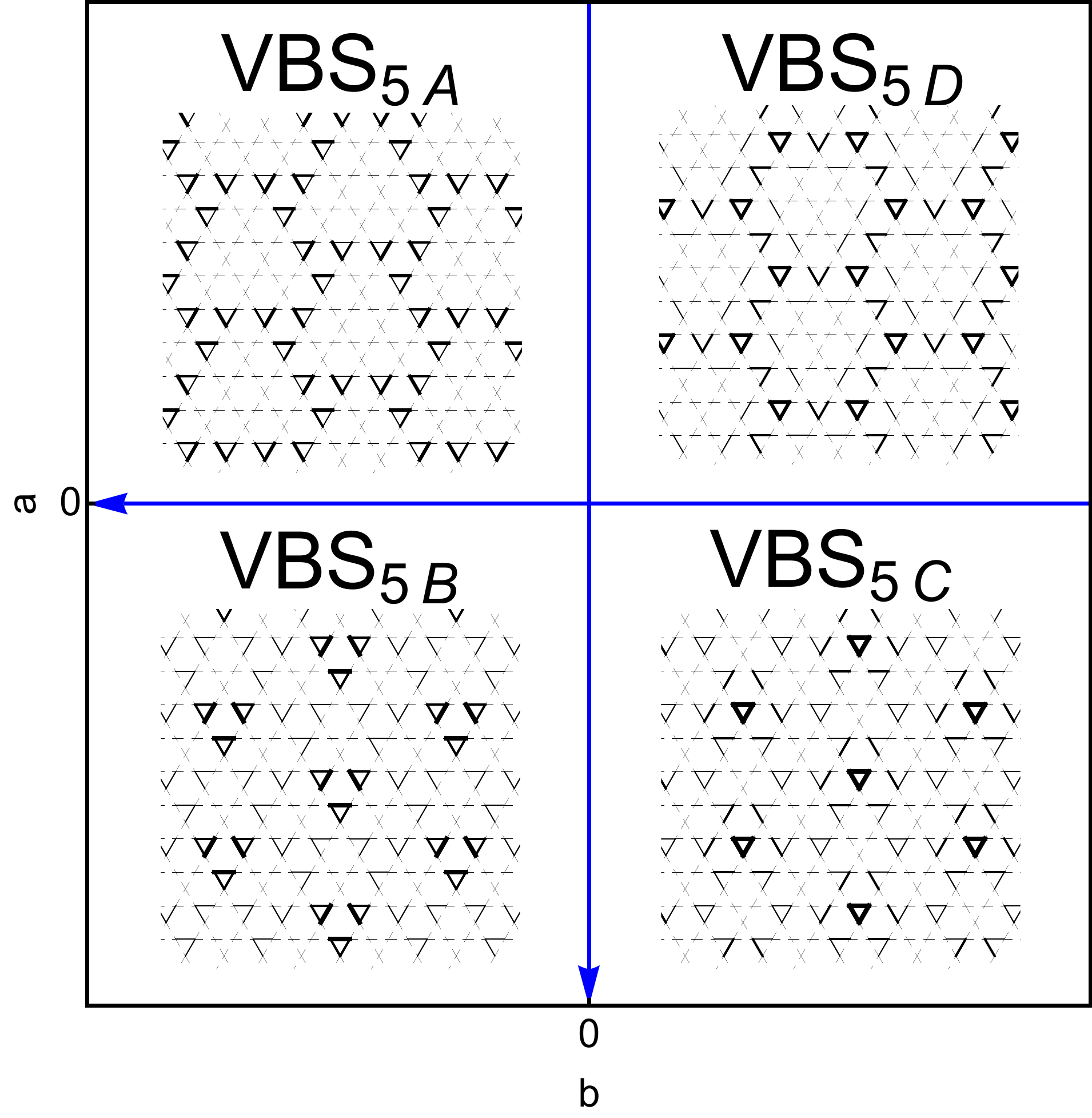}
\caption{(Color online) Mean-field phases diagram of the Landau theory [Eq.~(\ref{eq:GL-action})]. We have set $r = -1$, $u = 10$, and $v = 10$. The four phases are arranged to parallel Fig.~\ref{fig:case4-phases} by flipping the axes. While these do not satisfy the hard-core constraint, they tell us the symmetry breaking patterns that occur and is consistent with Fig.~\ref{fig:case4-phases}.}
\label{fig:phases}
\end{figure}

The fourth order Landau functional does not break continuous symmetries and the sixth order polynomials are necessary. 
Symmetry breaking patterns for vison condensation are depicted in Fig.~\ref{fig:phases}. When $J_2>J_1$, the interaction prefers visons at the centers of down triangles to those of up triangles, and up triangles have zero weight in all of the soft modes. This is allowed because the hard core dimer constraint has been relaxed. 
However we can still study the symmetry breaking patterns. As shown in Fig.~\ref{fig:phases}, the four distinct phases break the same symmetries that were broken in Fig.~\ref{fig:case4-phases} and are equivalent phases. 

Alternatively, we can include a further neighbor interaction that is only allowed in the case with anisotropy. This is shown in Fig.~\ref{fig:dice} on the right. This interaction has the same symmetries as the $A$ interactions in Eqs.~(\ref{eq:Hamiltonian}) and (\ref{eq:dual_Ising}). This gives the same result as above.




\begin{thebibliography}{99}
\bibitem{comment}
At the mean-field level, the dimer density deviation, $d-\bar{d}$, is computed by using the expectation value, $\langle \sigma_{ij}^x \rangle = \lambda_{\alpha\beta} \langle \phi_{\alpha} \rangle \langle \phi_{\beta} \rangle$. Here, the mean value of the vison field, $\langle \phi_{\alpha} \rangle$, is obtained from the mean-field, $\Psi_{MF}$, and Eq. (\ref{eq:vison-field}). In the VBS$_{1C}$ phase, $\langle \phi_{\alpha} \rangle$ is zero at every other sites on the dual triangular lattice so that $\langle \sigma_{ij}^x \rangle = 0$ at every link on the honeycomb lattice. Consequently, the dimer density is uniform on the kagome lattice. This also applies to the VBS$_{2C}$ phase.
\end{thebibliography}
\end{document}